\newif\ifAnon\Anonfalse
\newif\ifDraft\Draftfalse

\documentclass[conference]{IEEEtran}
\IEEEoverridecommandlockouts
\PassOptionsToPackage{bookmarks,hidelinks}{hyperref}
\usepackage{tugraz_defaults}
\usepackage{xcolor,colortbl}
\usepackage{enumitem}
\usepackage{subcaption}
\usepackage{amsmath}
\usepackage{mathtools}
\usepackage{tcolorbox}
\tcbuselibrary{listings}

\usepackage{pdfpages}

\newcommand{\ToolName}{SysFlow\xspace}

\DeclareMathAlphabet\mathbfcal{OMS}{cmsy}{b}{n}
\newcommand{\COne}{$\mathbfcal{C}\mathbf{1}$\xspace}
\newcommand{\CTwo}{$\mathbfcal{C}\mathbf{2}$\xspace}
\newcommand{\CThree}{$\mathbfcal{C}\mathbf{3}$\xspace}

\begin{document}
\makeatletter
\newcommand{\linebreakand}{%
  \end{@IEEEauthorhalign}
  \hfill\mbox{}\par
  \mbox{}\hfill\hspace{-0.7cm}\begin{@IEEEauthorhalign}
}
\makeatother

\date{}

\title{SFIP: Coarse-Grained Syscall-Flow-Integrity Protection in Modern Systems}

\ifAnon
\author{Anonymous author(s)}
\else
\author{
\IEEEauthorblockN{Claudio Canella}
\IEEEauthorblockA{Graz University of Technology\\
claudio.canella@iaik.tugraz.at}
\and
\IEEEauthorblockN{Sebastian Dorn}
\IEEEauthorblockA{Graz University of Technology\\
sebastian.dorn@student.tugraz.at}
\and
\IEEEauthorblockN{Daniel Gruss}
\IEEEauthorblockA{Graz University of Technology\\
daniel.gruss@iaik.tugraz.at}
\and
\linebreakand
\IEEEauthorblockN{Michael Schwarz}
\IEEEauthorblockA{CISPA Helmholtz Center\\ for Information Security\\
michael.schwarz@cispa.de}
}
\fi

\maketitle

\begin{abstract}
Growing code bases of modern applications have led to a steady increase in the number of vulnerabilities.
Control-Flow Integrity (CFI) is one promising mitigation that is more and more widely deployed and prevents numerous exploits.
CFI focuses purely on one security domain.
That is, transitions between user space and kernel space are not protected by CFI.
Furthermore, if user space CFI is bypassed, the system and kernel interfaces remain unprotected, and an attacker can run arbitrary transitions.

In this paper, we introduce the concept of syscall-flow-integrity protection (SFIP) that complements the concept of CFI with integrity for user-kernel transitions.
Our proof-of-concept implementation relies on static analysis during compilation to automatically extract possible syscall transitions. 
An application can opt-in to SFIP by providing the extracted information to the kernel for runtime enforcement. 
The concept is built on three fully-automated pillars:
First, a syscall state machine, representing possible transitions according to a syscall digraph model.
Second, a syscall-origin mapping, which maps syscalls to the locations at which they can occur.
Third, an efficient enforcement of syscall-flow integrity in a modified Linux kernel.
In our evaluation, we show that SFIP can be applied to large scale applications with minimal slowdowns.
In a micro- and a macrobenchmark, it only introduces an overhead of \SI{13.1}{\percent} and \SI{1.8}{\percent}, respectively.
In terms of security, we discuss and demonstrate its effectiveness in preventing control-flow-hijacking attacks in real-world applications.
Finally, to highlight the reduction in attack surface, we perform an analysis of the state machines and syscall-origin mappings of several real-world applications.
On average, SFIP decreases the number of possible transitions by \SI{38.6}{\percent} compared to seccomp and \SI{90.9}{\percent} when no protection is applied.
\end{abstract}

\section{Introduction}\label{sec:intro}
Vulnerablities in modern applications can be exploited by an attacker to gain arbitrary code execution within the application~\cite{Szekeres2013sok}.
Subsequently, the attacker can exploit further vulnerabilities in the underlying system to elevate privileges~\cite{Kemerlis2014}.
Such attacks can be mitigated in either of these two stages: the stage where the attacker takes over control of a victim application~\cite{Szekeres2013sok,Cowan1998}, or the stage where the attacker exploits a bug in the system to elevate privileges~\cite{Kemerlis2015,Kemerlis2012}.
Both researchers and industry have focused on eliminating the first stage, where an attacker takes over control of a victim application, by reducing the density of vulnerabilities in software, \eg by enforcing memory safety~\cite{Szekeres2013sok,Cowan1998}.
The second line of defense, protecting the system, has also been studied extensively~\cite{Kemerlis2015,Kemerlis2012,Ge2016kernel,Spengler2013uderef}.
For instance, sandboxing is a technique that tries to limit the available resources of an application, reducing the remaining attack surface. 
Ideally, an application only has the bare minimum of resources, \eg syscalls, that are required to work correctly.

Control-flow integrity~\cite{Abadi2005CFI} (CFI) is a mitigation that limits control-flow transfers within an application to a set of pre-determined locations.
While CFI has demonstrated that it can prevent attacks, it is not infallible~\cite{Goktas2014COP}.
Once it has been circumvented, the underlying system and its interfaces are once again exposed to an attacker as CFI does not apply protection across security domains.

In the early 2000s, Wagner and Dean~\cite{Wagner2001static} proposed an automatic, static analysis approach that generates syscall digraphs, \ie a k-sequence~\cite{Forrest1996self} of consecutive syscalls of length $2$.
A runtime monitor validates whether a transition is possible from the previous syscall to the current one and raises an alarm if it is not.
The Secure Computing interface of Linux~\cite{Edge2015seccomp}, seccomp, simplifies the concept by only validating whether a syscall is allowed, but not whether it is allowed in the context of the previous one. 
In contrast to the work by Wagner and Dean~\cite{Wagner2001static}, seccomp acts as an enforcement tool instead of a simple monitoring system.
Hence, false positives are not acceptable, as they would terminate a benign application.
Thus, we ask the following questions in this paper:

\emph{Can the concept of CFI be applied to the user-kernel boundary? 
Can prior syscall-transition-based intrusion detection models, \eg digraph models~\cite{Wagner2001static}, be transformed into an enforcement mechanism without breaking modern applications?}

In this paper, we answer both questions in the affirmative.
We introduce the concept of syscall-flow-integrity protection (SFIP), complementing the concept of CFI with integrity for user-kernel transitions.
Our proof-of-concept implementation relies on static analysis during compilation to automatically extract possible syscall transitions. 
An application can opt-in to SFIP by providing the extracted information to the kernel for runtime enforcement. 
SFIP builds on three fully-automated pillars, a syscall state machine, a syscall-origin mapping, and an efficient SFIP enforcement in the kernel.

The \textbf{syscall state machine} represents possible transitions according to a syscall digraph model.
In contrast to Wagner and Dean's~\cite{Wagner2001static} runtime monitor, we rely on an efficient state machine expressed as an $N \times N$ matrix ($N$ is the number of provided syscalls), that scales even to large and complex applications.
We provide a compiler-based proof-of-concept implementation, called \textit{\ToolName}\footnote{\url{https://github.com/SFIP/SFIP}}, that generates such a state machine instead of individual sets of k-sequences.
For every available syscall, the state machine indicates to which other syscalls a transition is possible.
Our syscall state machine (\ie the modified digraph) has several advantages including faster lookups ($\mathcal{O}(1)$ instead of $\mathcal{O}(M)$ with $M$ being the number of possible $k$-sequences), easier construction, and less and constant memory overhead.

The \textbf{syscall-origin mapping} maps syscalls to the locations at which they can occur.
Syscall instructions in a program may be used to perform different syscalls, \ie a bijective mapping between code location and syscall number is not guaranteed.
We resolve the challenge of these non-bijective mappings with a mechanism propagating syscall information from the compiler frontend and backend to the linker, enabling the precise enforcement of syscalls and their origin.
During the transition check, we additionally check whether the current syscall originates from a location at which it is allowed to occur.
For this purpose, we extend our syscall state machine with a syscall-origin mapping that can be bijective or non-bijective, which we extract from the program.
Consequently, our approach eliminates syscall-based shellcode attacks and imposes additional constraints on the construction of ROP chains.

The \textbf{efficient enforcement} of syscall-flow integrity is implemented in the Linux kernel.
Instead of detection, \ie logging the intrusion and notifying a user as is the common task for intrusion-detection systems~\cite{Kemmerer2002intrusion}, we focus on enforcement.
Our proof-of-concept implementation places the syscall state machine and non-bijective syscall-origin mapping inside the Linux kernel.
This puts our enforcement on the same level as seccomp, which is also used to enforce the correct behavior of an application. However, detecting the set of allowed syscalls for seccomp is easier.
As such, our enforcement is an additional technique to sandbox an application, automatically limiting the post-exploitation impact of attacks.
We refer to our enforcement as \textit{coarse-grained syscall-flow-integrity protection}, effectively emulating the concept of control-flow integrity on the syscall level. %

We evaluate the performance of SFIP based on our reference implementation.
In a microbenchmark, we only observe an overhead on the syscall execution of up to \SI{13.1}{\percent}, outperforming seccomp-based protections.
In a macrobenchmark using nginx and ffmpeg, we observe an overhead of \SI{1.5}{\percent} and \SI{1.8}{\percent} compared to an unprotected version, respectively.
We evaluate the one-time overhead of extracting the information from a set of real-world applications.
In the worst case, we observe an increase in compilation time by factor \SIx{28}.

We evaluate the security of the concept of syscall-flow-integrity protection in a security analysis with special focus on control-flow hijacking attacks.
We evaluate our approach on real-world applications in terms of number of states (\ie syscalls with at least one outgoing transition), number of average transitions per state, and other security-relevant metrics.
Based on this analysis, SFIP, on average, decreases the number of possible transitions by \SI{38.6}{\percent} compared to seccomp and \SI{90.9}{\percent} when no protection is applied.
Against control-flow hijacking attacks, we find that in nginx, a specific syscall can, on average, only be performed at the location of \SIx{3} syscall instructions instead of in \SIx{318} locations.
We conclude that syscall flow integrity increases system security substantially while only introducing acceptable overheads.

To summarize, we make the following contributions:
\begin{compactenum}
  \item We introduce the concept of (coarse-grained) \textit{syscall-flow-integrity protection} (SFIP) to enforce legitimate user-to-kernel transitions based on static analysis of applications.
  \item Our proof-of-concept SFIP implementation is based on a syscall state machine and a mechanism to validate a syscall's origin.
  \item We evaluate the security of SFIP quantitatively, showing that the number of possible syscall transitions is reduced by \SI{90.9}{\percent} on average in a set of \SIx{6} real-world applications, and qualitatively, by analyzing the implications of SFIP on a real-world exploit.
  \item We evaluate the performance of our SFIP proof-of-concept implementation, showing an overhead of \SI{13.1}{\percent} in a microbenchmark and \SI{1.8}{\percent} in a macrobenchmark.
\end{compactenum}

\textbf{Outline}
\Cref{sec:background} provides background.
\Cref{sec:design} then provides our threat model, discusses the high-level concept of syscall-flow integrity, and discusses the challenges of extracting the information required for SFIP and enforcing it.
\cref{sec:implementation} provides implementation details, and \cref{sec:eval} evaluates it with regard to performance and security.
\cref{sec:discussion} discusses limitations as well as future and related work.
\cref{sec:conclusion} then concludes the work.

\section{Background}\label{sec:background}
This section discusses the necessary background for this work.

\subsection{Sandboxing}
Sandboxing is a technique that tries to constrain the resources of an application to the absolute minimum necessary for the application to still work correctly.
For instance, a sandbox might limit an application's access to files, to the network, or the syscalls it can perform.
As such, a sandbox is often the last line of defense in an already exploited application, trying to limit the post-exploitation impact.
Nowadays, sandboxes are widely deployed in various applications, including in mobile operating systems~\cite{Android2017seccomp,AndroidAppSandbox} and browsers~\cite{Firefox2019sandbox,Reis2019SiteIsolation,Firefox2019fission}.
Linux itself also provides various methods for sandboxing, including SELinux~\cite{SELinuxFAQ}, AppArmor~\cite{AppArmor}, or seccomp~\cite{Edge2015seccomp}.

\subsection{Digraph Model}
The behavior of an application can be modeled by the sequence of syscalls it performs.
In intrusion-detection systems, windows of consecutive syscalls, so-called \textit{k-sequences}, have been used~\cite{Forrest1996self}.
A special case of k-sequences are sequences of length $k=2$, which are commonly referred to as digraphs~\cite{Wagner2001static}.
A model built upon these digraphs can allow for easier construction and more efficient checking while reducing the accuracy in the detection~\cite{Wagner2001static}.
That is because only the previous syscall and the current one are considered, instead of a longer sequence.

\subsection{Linux Seccomp}
The syscall interface is a security-critical interface that the Linux kernel exposes to userspace applications.
Applications rely on the syscall interface to request the execution of privileged tasks from the kernel.
Hence, securing this interface is crucial to improving the overall security of the system.

To better secure this interface, the kernel provides Linux Secure Computing (seccomp).
A benign application first creates a filter that contains all the syscalls it intends to perform over its lifetime and then passes this filter to the kernel.
Upon a syscall, the kernel checks whether the executed syscall is part of the set of syscalls defined in the filter and either allows or denies it.
As such, seccomp can be seen as a k-sequence of length $1$.
In addition to the syscall itself, seccomp can filter static syscall arguments. 
Hence, seccomp is an essential technique to limit the post-exploitation impact of an exploit as unrestricted access to the syscall interface allows an attacker to arbitrarily read, write, and execute files.
An even worse case is when the syscall interface itself is exploitable, as this can lead to privilege escalation~\cite{Kemerlis2014,Kemerlis2015,Kemerlis2012}.

\subsection{Runtime Attacks}
One of the root causes for successful exploits are memory safety violations.
One typical variant of such a violation are buffer overflows, enabling an attacker to modify the application in a malicious way~\cite{Szekeres2013sok}.
An attacker tries to use such a buffer overflow to overwrite a code pointer, such that the control flow can be diverted to an attacker-chosen location, \eg to previously injected \textit{shellcode}.
Attacks relying on shellcode have become harder to execute on modern systems due to data normally not being executable~\cite{Szekeres2013sok,Microsoft2021dep}.
Therefore, attacks have to rely on already present, executable code parts, so-called \textit{gadgets}. 
These gadgets are chained together to perform an arbitrary attacker-chosen task~\cite{Nergal2001ret2libc}.
Shacham further generalized this attack technique as return-oriented programming (ROP)~\cite{Shacham2007}.
Similar to control-flow-hijacking attacks that overwrite pointers~\cite{Shacham2007,Checkoway2010JOP,Lan2015loop,Goktas2014COP,Schuster2015COOP}, memory safety violations can also be abused in data-only attacks~\cite{Rogowski2017,Ispoglou2018}.

\subsection{Control-Flow Integrity}
Control-flow integrity~\cite{Abadi2005CFI} (CFI) is a concept that restricts an application's control flow to valid execution traces, \ie it restricts the targets of control-flow transfer instructions.
This is enforced at runtime by comparing the current state of the application to a set of pre-computed states.
Control-flow transfers can be divided into forward-edge and backward-edge transfers~\cite{Burow2017cfi}.
Forward-edge transfers transfer control flow to a new destination, such as the target of an (indirect) jump or call.
Backward-edge transfers transfer the control flow back to a location that was previously used in a forward edge, \eg a return from a call.
Furthermore, CFI can be subdivided into coarse-grained and fine-grained CFI.
In contrast to fine-grained CFI, coarse-grained CFI allows for a more relaxed control-flow graph, allowing more targets than necessary~\cite{Davi2014coarsecfi}.

\section{Design of Syscall-Flow-Integrity Protection}\label{sec:design}

In this section, we define the threat model for SFIP (\Cref{sec:threat-model}), the high-level design (\Cref{sec:high-level}), and the challenges for such an approach (\Cref{sec:challenges}). 

\subsection{Threat Model}\label{sec:threat-model}
SFIP is applied to userspace applications. 
We assume that the protected application is benign but potentially contains a vulnerability that allows an attacker to execute arbitrary code within the application.
The post-exploitation then targets the operating system through the syscall interface to gain kernel privileges.
SFIP restricts the execution of syscalls in two ways.
First, a syscall is allowed if the state machine contains a valid transition from the previous syscall to the current one.
Second, a syscall is allowed if it is a valid entry in the syscall-origin mapping, \ie it originates from a pre-determined location.
If either one is violated, the application is terminated by the kernel.
Similar to prior work~\cite{Canella2021chestnut,Ghavamnia2020temporal,DeMarinis2020,Ghavamnia2020confine}, our protection is orthogonal but fully compatible with defenses such as CFI, ASLR, NX, or canary-based protections.
Therefore, the security it provides to the system remains even if these other protections have been circumvented.
Side-channel and fault attacks~\cite{Kim2014,Yarom2014Flush,Kocher2019,Lipp2018meltdown,VanSchaik2019RIDL,Schwarz2019ZL} on the state machine or syscall-origin mapping are out of scope.

\begin{figure}
  \centering
  \resizebox{0.9\hsize}{!}{
    \includegraphics{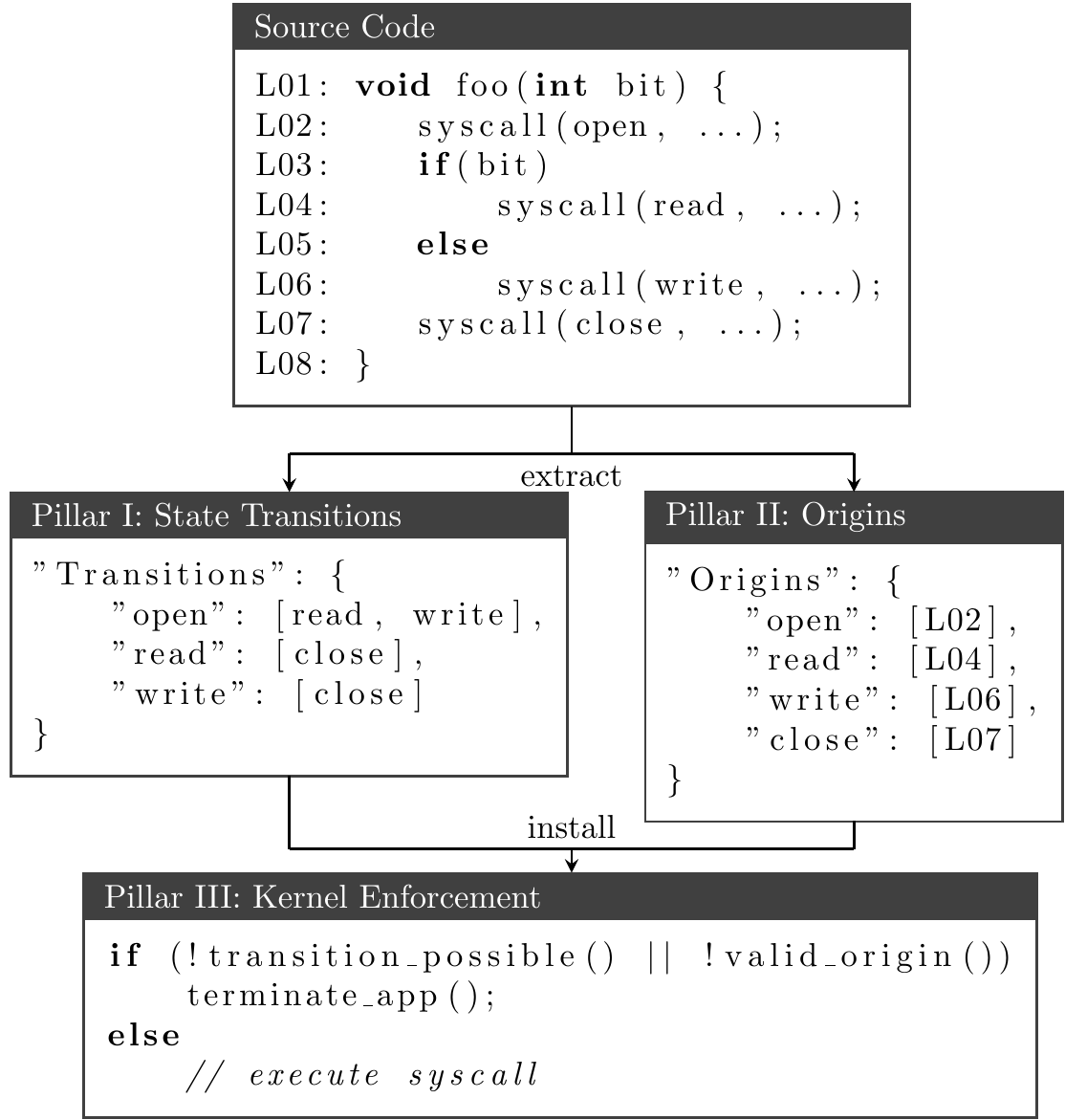}
  }
  \caption{The three pillars of SFIP on the example of a function.
  The first pillar models possible syscall transitions, the second maps syscalls to their origin, and the third enforces them.}
  \label{fig:high-level-design}
\end{figure}

\subsection{High-Level Design}\label{sec:high-level}
In this section, we discuss the high-level design behind SFIP.
Our approach is based on three pillars: a digraph model for syscall sequences, a per-syscall model of syscall origin, and the strict enforcement of these models. 
\Cref{fig:high-level-design} illustrates this high-level design.

For our first pillar, we rely on the idea of a digraph model from Wagner and Dean~\cite{Wagner2001static}. %
Digraphs are a special case of the \textit{k-sequences}, \ie windowed sequences of consecutive syscalls, introduced by Forrest~\etal\cite{Forrest1996self}.
They have been proposed in the early 2000s but neither the generation nor the runtime monitoring has been evaluated in the context of modern, large-scale applications. %
In a digraph model~\cite{Wagner2001static}, the sequence length is fixed to \SIx{2}. %
For our sycall-flow-integrity protection, we also use a sequence length of \SIx{2}, but rely on a more efficient construction and in-memory representation. %
In contrast to their approach, we express the set of possible transitions not as individual k-sequences, but as a global syscall matrix of size $N \times N$, with $N$ being the number of available syscalls.
We refer to the matrix as our \textit{syscall state machine}.
With this representation, verifying whether a transition is possible is a simple lookup in the row indicated by the previous syscall and the column indicated by the currently executing syscall.
Even though the representation of the sequences differs, the set of valid transitions remains the same: every transition that is marked as valid in our syscall state machine must also be a valid transition if expressed in the way discussed by Wagner and Dean.
The reason is that both are generated from the source code of the application; hence, any transition valid in one representation must be valid in the other.
Our representation has several advantages though, that we explore in this paper, namely faster lookups ($\mathcal{O}(1)$), less memory overhead, and easier construction.

Our syscall state machine can already be used for coarse-grained SFIP to improve the security of the system (\cf \cref{sec:eval-security}). 
However, the second pillar, the validation of the origin of a specific syscall, further improves the provided security guarantees by adding additional, enforcable information.
The basis for this augmentation is the ability to precisely map syscalls to the location at which they can be invoked, independent of whether it is a bijective or non-bijective mapping.
We refer to the resulting mapping as our \textit{syscall-origin mapping}.
For instance, our mapping might contain the information that the syscall instruction located at address \texttt{0x7ffff7ecbc10} can only execute the syscalls \textit{write} and \textit{read}.
By design, this concept alone eliminates shellcode attacks:
Neither unaligned execution (\eg in a ROP chain) nor code inserted at runtime is in our syscall-origin mapping.
Thus, syscalls can only be executed at already existing syscall instructions.

The third pillar of SFIP is the enforcement of the syscall state machine and the syscall-origin mapping.
Wagner and Dean~\cite{Wagner2001static} proposed their runtime monitoring as a concept for intrusion-detection systems, which are designed to detect and log violations. 
There is still a domain expert involved to decide any further action~\cite{Kemmerer2002intrusion}.
The difference between monitoring and enforcement is that enforcement cannot afford false positives as this immediately leads to the termination of the application in benign scenarios.
On the other hand, enforcement provides better security than monitoring as immediate action is undertaken, completely eliminating the time window for a possible exploit.
Thus, by the use case of SFIP, namely enforcement of syscall-flow integrity, our concept is more closely related to seccomp but harder to realize than seccomp-based enforcement of syscalls.

\subsection{Challenges}\label{sec:challenges}
Previous automation work outlined several challenges that need to be solved for automatically detecting the syscalls an application uses so that seccomp can correctly filter them~\cite{Canella2021chestnut}.
We investigated several works~\cite{Canella2021chestnut,DeMarinis2020,Ghavamnia2020temporal} that propose automated seccomp-filter generation and find that each solves these challenges, but none provides the full information required for SFIP.
Hence, we identify challenges that must be solved by each proposed approach to provide the information required for SFIP.
Due to the switch from runtime monitoring to enforcement, the generation of our syscall state machine and syscall-origin mapping must be precise as false positives immediately lead to the termination of the application.
Therefore, the challenges primarily focus on precise syscall information and inter- and intra-procedural control-flow transfer information.
We illustrate the challenges using a simple dummy program shown in \Cref{lst:cfg-example}. 

\begin{listing}[t]
  \begin{lstlisting}[language=C,style=customc]
void foo(int bit, int nr) {
  syscall(open, ...);
  if(bit)
    syscall(read, ...);
  else
    syscall(nr, ...);
  bar(...);
  syscall(close, ...);
}
 \end{lstlisting}
 \caption{Example of a dummy program with multiple syscall-flow paths.}
 \label{lst:cfg-example}
 \end{listing}

\paragraph{\COne: Precise Per-Function Syscall Information}
The first challenge focuses on precise per-function syscall information.
This challenge must be solved for the generation of the syscall state machine as well as the sycall-origin map, although for different reasons.
For seccomp-based approaches, \ie k-sequence of length \SIx{1}, an automatic approach only needs to identify the set of syscalls within a function, \ie the exact location of the syscalls is irrelevant.
In our example, seccomp only requires the information that our function \texttt{foo} contains the syscalls \textit{open}, \textit{read}, and \textit{close} and the one identified by the function parameter \textit{nr}.
This does not hold for SFIP, which requires precise information at which location a specific syscall is executed.
Thus, we have to detect that the first syscall instruction always executes the \textit{open} syscall, the second executes \textit{read}, and the third syscall instruction can execute any syscall that can be specified via \textit{nr}.
For the state machine generation, the precise information of syscall locations provides parts of the information required to correctly generate the sequence of syscalls.
For the syscall-origin map, the precise information allows generating the mapping of syscall instructions to actual syscalls in the case where syscall numbers are specified as a constant at the time of invocation.

\paragraph{\CTwo: Argument-based Syscall Invocations}
The second challenge extends upon \COne as it concerns syscall locations where the actual syscall executed cannot be easily determined at the time of compilation.
When parsing the function \texttt{foo}, we can identify the syscall number for all invocations of the \texttt{syscall} function where the number is specified as a constant.
The exception is the third invocation, as the number is provided by the caller of the \texttt{foo} function.
As the call to the function, and hence the actual syscall number, is in a different translation unit than the actual syscall invocation, the possibility for a non-bijective mapping exists.
Still, an automated approach must be able to determine all possible syscalls that can be invoked at each syscall instruction.

\paragraph{\CThree: Correct Inter- and Intra-Procedural Control-Flow Graph}
Precise per-function syscall information on its own is not sufficient to generate syscall state machines due to the non-linearity of typical code.
Solving \COne and \CTwo provides us with the information which syscalls occur at which syscall location, but does not provide us with information in which sequence they can be executed in.
A trivial construction algorithm can simply assume that each syscall within a function can follow each other syscall within the function, but this overapproximation leads to imprecise state machines.
Such an approach accepts a transition from \textit{read} to the syscall identified by \textit{nr} as valid, even though it cannot occur within our example function.

Therefore, we need to determine the correct inter- and intra-procedural control-flow transfers in an application.
The correct intra-procedural control-flow graph allows determining the possible sequences within a function.
In our example, and if function \texttt{bar} does not contain any syscalls, it provides us with the information that the sequence of syscalls \textit{open $\to$ read $\to$ close} is valid, while a sequence of \textit{open $\to$ nr $\to$ close} (where \textit{nr} $\neq$ \textit{read}) is not.

Even in the presence of a correct intra-procedural control-flow graph, we cannot reconstruct the syscall state machine of an application as information is missing on the sequence of syscalls from other called functions.
For instance, if function \texttt{bar} contains at least one syscall, the sequence of \textit{open $\to$ read $\to$ close} is no longer valid.
Hence, we additionally need to recover the precise location where control flow is transferred to another function, as well as the target of this control-flow transfer.
By combining the inter- and intra-procedural control-flow graph, the correct syscall sequences of an application can be modeled.

Constructing a precise control-flow graph is known to be a hard task to solve efficiently~\cite{Andersen1994points,Hind2001pointer}, especially in the presence of indirect control-flow transfers.
These algorithms are often cubic in the size of the application, which makes them infeasible for large-scale applications.
In the construction of the control-flow graph and, by extension the generation of the syscall state machine and syscall-origin mapping, other factors, such as aliased and referenced functions must be considered as well as functions that are passed as arguments to other functions, \eg the entry function for a new thread created with \texttt{pthread\_create}.
Any form of imprecision can lead to the termination of the application by the runtime enforcement.

\section{Implementation}\label{sec:implementation}
In this section, we discuss our proof-of-concept implementation \ToolName and how we systematically solve the challenges outlined in \cref{sec:challenges} to provide fully automated SFIP.
We discuss all three pillars of SFIP.
First, we discuss our implementation to extract our syscall state machine. %
Second, we discuss our implementation to extract a syscall-origin mapping, which augments the concept of syscall-sequence checks.
Third, we discuss our modified kernel that enforces the application's behavior instead of simply monitoring it.
We also discuss our support library, which is required for setting up the enforcement, similar to \textit{libseccomp}.

\paragraph{\ToolName}
\ToolName automatically generates the state machine and the syscall-origin mapping while compiling an application.
As the basis of \ToolName we considered the works by Ghavamnia~\etal\cite{Ghavamnia2020temporal} and Canella~\etal\cite{Canella2021chestnut}.
Both tools are already capable of extracting most syscall numbers and a reasonably-precise call graph, although only for seccomp-based protection.
However, neither of them solves the challenges we identified in \cref{sec:challenges}, making both equally well suited as a basis for our work.
As Ghavamnia~\etal\cite{Ghavamnia2020temporal} report significantly higher extraction times, we opted for the work by Canella~\etal\cite{Canella2021chestnut} as a basis.
Consequently, \ToolName is also built on top of LLVM 10.

\subsection{State-Machine Extraction}

\begin{figure}
  \centering
  \resizebox{1\hsize}{!}{
    \includegraphics{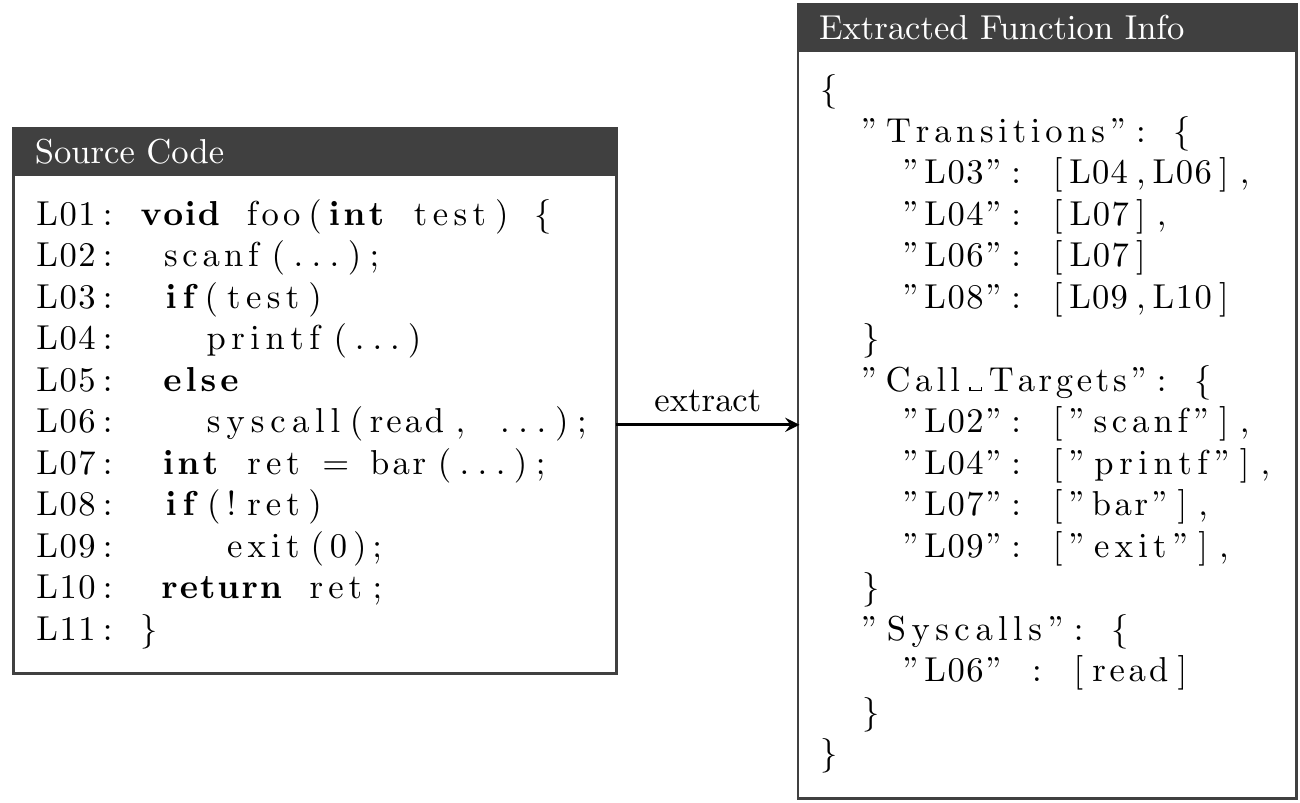}
  }
  \caption{A simplified example of the information that is extracted from a function.
  \textit{Transitions} identifies control-flow transfers between basic blocks, \textit{Call Targets} the location of a call to another function and the targets name, \textit{Syscalls} the location of the syscall and the corresponding syscall number.}
  \label{fig:compiler-control-flow-info}
\end{figure}

In \ToolName, the linker is responsible for creating the final state machine.
The construction works as follows:
The linker starts at the main function, \ie the user-defined entry point of an application, and recursively follows the ordered set of control-flow transfers.
Initialization functions, \eg musl's \textit{\_\_libc\_start\_main}, are not analyzed, as the enforcement is not activated before the \textit{main} function.
Upon encountering a syscall location, the linker adds a transition from the previous syscall(s) to the newly encountered syscall.
If control flow continues at a different function, the set of last valid syscall states is passed to the recursive visit of the encountered function.
Upon returning from a recursive visit, the linker updates the set of last valid syscall states and continues processing the function.
During the recursive processing, it also considers aliased and referenced functions.

A special case, and source of overapproximation, are indirect calls, which we address with appropriate techniques from previous works~\cite{Canella2021chestnut,DeMarinis2020,Ghavamnia2020confine}.
At the site of the indirect call, we know the signature of the function that is indirectly called.
We compare this signature with the signature of all functions that have their address taken.
All functions that match are then processed in the same way as above.

When the linker reaches the end of the \textit{main} function, it has processed all reachable functions.
In the process, it has generated the set of valid transitions for each syscall.
The resulting syscall state machine is then embedded in the static binary.
Our support library is automatically included in the binary as well and is responsible for installing the state machine when the application is launched.
We discuss the support library in more detail in \cref{sec:implementation-support}.

The process of building the state machine requires that precise information of the syscalls a function executes (\COne) and a control-flow graph of the application (\CThree) is available to the linker.
Both the front- and backend are involved in collecting this information.
The frontend extracts the information from the LLVM IR generated from C source code, while the backend extracts the information from assembly files.
To propagate the information to the linker, both the front- and backend store it in the resulting object file.
\cref{fig:compiler-control-flow-info} illustrates the information that is extracted from a function.

\paragraph{Extracting Precise Syscall Information}
In the frontend, we iterate over every IR instruction of a function and determine whether it is an inline assembly syscall instruction or a call to one of the libc syscall wrappers.
In most cases, the syscall number is specified as a constant, allowing the compiler to easily associate it with the location of the syscall.
In the backend, we iterate over every assembly instruction of a function specified in an assembly file.
On x86, the syscall number is placed in the \texttt{RAX} register before the syscall instruction.
Hence, we track the value that is moved into the register.
Once we encounter a syscall instruction, we associate the instruction with the value in \texttt{RAX}.
Extracting the information in the front- and backend successfully solves \COne.

\paragraph{Extracting Precise Control-Flow Information}
Recovering the control-flow graph (\CThree) in the frontend requires two different sources of information: IR call instructions and successors of basic blocks.
The former allows tracking inter-procedural control-flow transfers while the latter allows tracking intra-procedural transfers.
For inter-procedural transfers, we iterate over every IR instruction and determine whether it is a call to an external function.
For direct calls, we store the target of the call; for indirect calls, we store the function signature of the target function.
In addition, we also gather information on referenced and aliased functions, as well as functions that are passed as arguments to other functions.
For the intra-procedural transfers, we track the successors of each basic block.
In the backend, we perform similar steps, although on a platform-specific assembly level instead of the IR level.
Extracting this information in the front- and backend successfully solves \CThree.

\subsection{Syscall-Origin Extraction}

In \ToolName, the linker also generates the final syscall-origin mapping.
The mapping maps all reachable syscalls to the locations at which they can occur, relative to the start of the function that encapsulates them.
We extract the information as an offset instead of an absolute position to facilitate compatibility with ASLR.
The linker identifies all functions reachable via the \textit{main} function and adds their syscall map to the mapping.
The resulting mapping is embedded in the final static binary from where our support library can extract it to make the final address computations (\cf \cref{sec:implementation-support}).

The linker requires precise information of syscalls, \ie their offset relative to the start of the encapsulating function, and a precise call graph of the application.
Both the front- and backend are responsible for providing this information.
\cref{fig:sysloc-extraction} illustrates the extraction.
From the frontend, the syscall information generated by the state machine extraction is re-used (\COne); hence, we do not discuss it again but focus on the additional changes in the backend.
Extracting the offset is a complex process that requires adding and propagating partial information through various parts of the backend. %
The main reason for this complex process is the non-fixed instruction size on x86. %
An additional problem is the possibility of non-bijective syscall mappings, which must also be resolved (\CTwo).

\paragraph{Non-Fixed Instruction Size}
On architectures with fixed instruction sizes, computing the offset of an instruction relative to the start of the function is trivial.
On x86, with its non-fixed instruction size, the final instruction size is only known once relocations have been decided, but at this point, no information on syscalls is available.
Hence, earlier stages must add and propagate syscall information to make it available.

\paragraph{Non-Bijective Syscall Mappings}
If the syscall number cannot be determined at the location of a syscall instruction, a non-bijective mapping exists for the instruction, \ie multiple syscalls can be executed through it.
An example of such a case is shown in \cref{lst:cfg-example}.
Instances of this are non-inlined calls to the syscall wrapper functions provided by libc, \ie \texttt{syscall()} or \texttt{syscall\_cp()}, as syscall number and syscall instruction are in different translation units.
In such cases, the backend itself is not able to create a mapping of a syscall to the syscall instruction.
Hence, it must propagate the syscall number and the syscall offset from their respective translation unit to the linker, which can then merge it.
This allows the linker to finally solve \CTwo.

\begin{figure}
  \centering
  \resizebox{1\hsize}{!}{
    \includegraphics{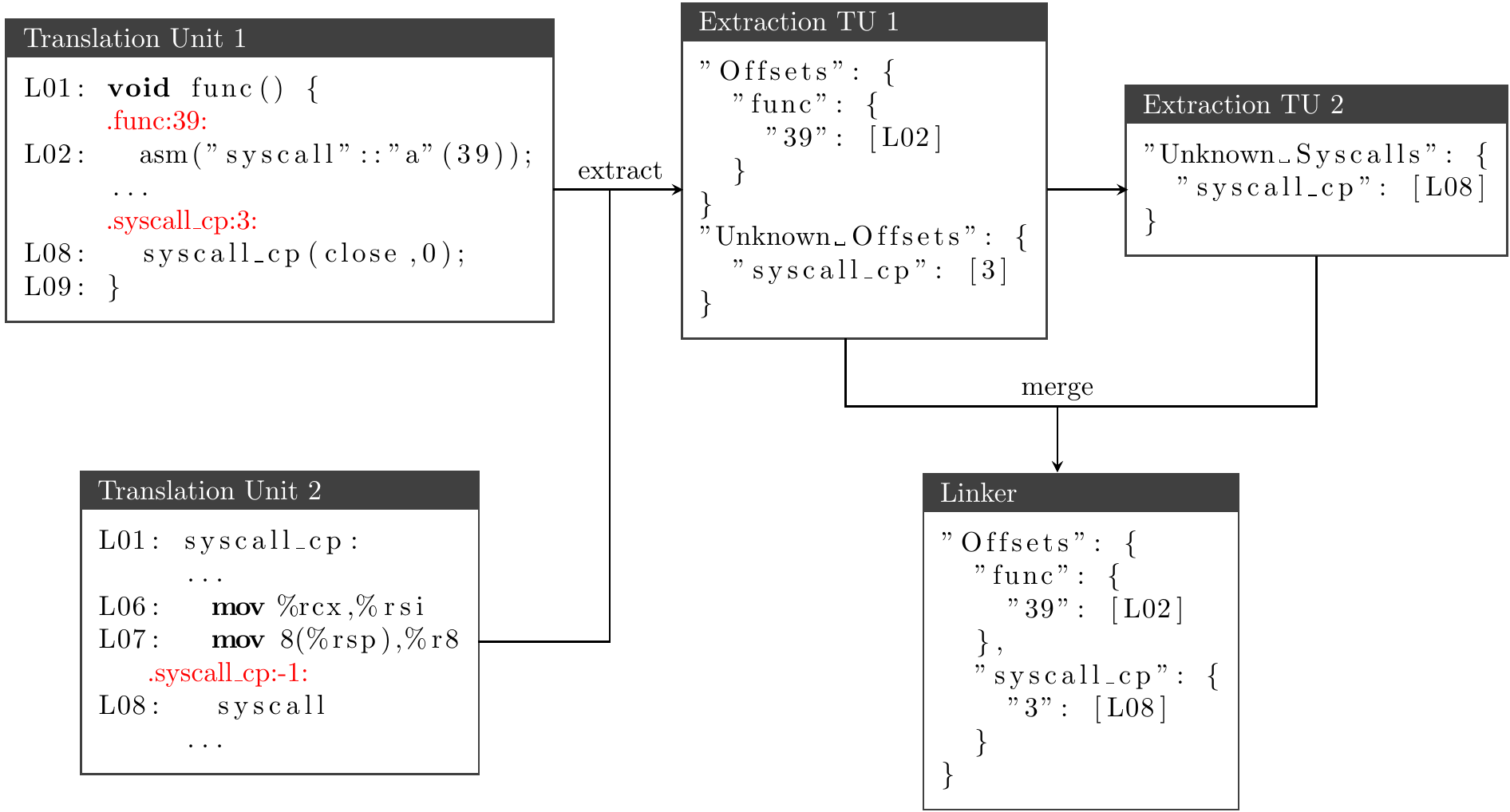}
  }
  \caption{A simplified example of the syscall-origin extraction.
  We insert labels (red) that mark the location of each syscall and encode available information for it.
  In the extraction, we deconstruct the label and calculate the offset using the label's address from the symbol table.
  The linker combines the information from each translation unit and generates the final syscall-origin mapping.}
  \label{fig:sysloc-extraction}
\end{figure}

\paragraph{Propagating Syscall Information}
A new pass in the backend iterates over all functions in the translation unit and the instructions that comprise them.
For every syscall that is encountered, the pass emits a label before it.
The label is emitted independent of whether the syscall is performed using an inlined syscall instruction or a call to one of the wrappers.
The label serves two purposes: it is used to propagate necessary information through the backend, and enables the computation of the syscall offset.
For an inlined syscall, we embed the name of the encapsulating function and, if available, the syscall number in the label.
For a call to one of the syscall wrappers, we embed the name of the wrapper and, if available, the syscall number.
If the syscall number is not available (\CTwo), we embed the value \SIx{-1}.
This identifies a syscall where the linker is responsible for constructing the correct function-to-offset-to-syscall mapping.

In addition to the new backend pass, we extend the AsmParser to emit labels for syscalls originating in assembly files.
This is necessary since our MachineFunctionPass cannot add these labels as it iterates over the machine IR generated from the LLVM IR.

\paragraph{Offset Calculation}
Before the object file is generated, the backend determines relocations, which determines final instruction sizes, and computes the symbol table.
The latter assigns an address to the injected labels. %
With the symbol table computed, all information required to create the translation unit's syscall mapping is available.
The symbol table provides the offset, and the label's content provides the remaining information required to create the mapping.
If the label contains the syscall number, we can directly create the mapping of function to syscall number to syscall offset, \eg \texttt{malloc} performs a \textit{futex} syscall at offset $0x209$.
If the label contains \SIx{-1}, we only create a mapping of function to offset, \eg \texttt{syscall\_cp} performs an unknown syscall at offset $0x1c$
If the label contains the name of a syscall wrapper, we create a mapping of wrapper function to syscall number, \eg \texttt{syscall\_cp} performs syscall \textit{read} at an unknown offset.
The linker merges the latter two once it has extracted the information from all object files and solves \CTwo.
This finally allows the linker to create the complete syscall-origin mapping of the application.

\subsection{Installation}\label{sec:implementation-support}
In this section, we discuss the implementation of our support library, which is responsible for extracting the generated information from the binary and installing it in the kernel.
\ToolName automatically adds the library to the static binary during compilation.
It contains a constructor that is run before the execution starts \textit{main}, similar to what has been done in previous work~\cite{Canella2021chestnut,DeMarinis2020}.

\paragraph{State Machine}
For each syscall, the binary contains a list of all other reachable syscalls.
The library converts this information into an $N \times N$ matrix, \ie the state machine, with $N$ being the number of syscalls available.
Valid transitions are indicated by a \SIx{1} in the matrix, invalid ones with a \SIx{0}.
This design allows for fast checks and constant memory overhead, independent of the number of possible transitions.
The resulting state machine is sent to the kernel and installed.

\paragraph{Syscall Origins}
Installing the syscall-origin information requires additional pre-processing to determine the final location of every syscall.
The support library extracts the symbol table of the static binary to retrieve the load address of every function.
If a function contains a syscall, the offset of the syscall is added to the load address of the function.
The corresponding entry in the syscall-origin mapping is updated with the result.
The final syscall-origin mapping is sent to the kernel and installed.

\subsection{Kernel Enforcement}\label{sec:implementation-kernel}
In this section, we discuss the third and final pillar of SFIP: enforcement of the syscall flow and origin.
As we previously discussed, SFIP does not perform simple runtime monitoring as is common in intrusion-detection systems, but runtime enforcement.
Hence, every violation leads to immediate process termination.

Our kernel is based on Linux kernel version 5.13 configured for Ubuntu 21.04.
We first discuss the common parts of both the state machine as well as the syscall origins, namely the following three modifications of the kernel:

First, a new syscall, \textit{SYS\_syscall\_sequence}, which takes as arguments the state machine, the syscall-origin mapping, and a flag that identifies the requested mode, \ie is state-machine enforcement requested, syscall-origin enforcement, or both.
This is necessary to make the kernel aware of our syscall-flow integrity information.
The kernel copies the corresponding data and stores it in the \textit{task\_struct} of the current process.
It rejects updates to already installed syscall-flow information.
The kernel also sets the \textit{NO\_NEW\_PRIVS} flag, similar to seccomp.
Consequently, an unprivileged process cannot apply a malicious state machine or syscall origins before invoking a setuid binary or other privileged programs using one of the \textit{exec} syscalls~\cite{Edge2012nonewprivs}.

Second, we follow the example of seccomp and modify the kernel so that our syscall-flow integrity checks are executed before every syscall if the process has requested them.
For this purpose, we create a new \textit{syscall\_work\_bit} entry, which determines whether or not the kernel uses the slow syscall entry path, like in seccomp, to ensure that our checks are executed.
Upon installation, we set the respective bit in the \textit{syscall\_work} flag in the \textit{thread\_info} struct of the requesting task.

Third, the syscall-flow information has to be stored and cleaned up properly.
As it is never modified after installation, it can be shared between the parent and child processes and threads.
Hence, only a reference count for the stored information is required.
The current state, \ie the previously executed syscall, is not shared between threads or processes.
Thus, it is necessary to modify the \textit{copy\_process} function to copy the reference and the initial current state into every new process and thread if the parent has it installed.
For cleanup, we modify the \textit{release\_task} function that is called upon task cleanup.
There, we decrease the reference counter, and if it reaches 0, we free the respective memory.

\paragraph{Enforcing State Machine Transitions}
Overall, the process of enforcing our state machine is very efficient.
Each thread and process tracks its own current state in the state machine, which is laid out as a flattened $N \times N$ matrix.
As we enforce sequence lengths of size $2$, storing the previously executed syscall as the current state is sufficient for the enforcement.
Due to the design of our state machine, verifying whether a syscall is allowed is a single lookup in the matrix at the location indicated by the previous and current syscall.
If the entry indicates a valid transition, we update our current state to the currently executing syscall and continue with the syscall execution.
If the entry does not indicate a valid transition, the application tries to perform a syscall that it should not be executing.
The kernel immediately terminates the offending application.
The simple state machine lookup, with a complexity of $\mathcal{O}(1)$, ensures that only a small overhead is introduced to the syscall (\cf \cref{sec:microbenchmark,sec:macrobenchmark}).

\paragraph{Enforcing Syscall Origins}
The enforcement of the syscall origins is also very efficient due to its design.
As previously discussed, our syscall-origin mapping maps syscall numbers to their respective virtual addresses.
Hence, our modified kernel uses the current syscall to retrieve the set of possible locations from the mapping. %
It then checks whether the current RIP, minus the size of the syscall instruction itself, is a part of the retrieved set.
If so, the syscall originates from a valid location, and we continue its execution.
Otherwise, the application requested the syscall from an unknown location, which results in the kernel immediately terminating it.
By design, the complexity of this lookup is $\mathcal{O}(N)$, with $N$ being the number of valid offsets for that syscall.
We evaluate typical values of $N$ in \cref{sec:sysloc-analysis}.

\section{Evaluation}\label{sec:eval}
In this section, we evaluate the general idea of SFIP and our proof-of-concept implementation \ToolName.
In the evaluation, we focus on the performance and security of the syscall state machines and syscall-origins individually, and combined.
We evaluate the overhead introduced on syscall executions in both a micro- and macrobenchmark.
We also evaluate the time required to extract the required information from a selection of real-world applications.

Our second focus is the security provided by SFIP.
We first consider the protection SFIP provides against control-flow hijacking attacks.
We evaluate the security of pure syscall-flow protection, pure syscall-origin protection, and combined protection.
We then discuss mimicry attacks and how SFIP makes such attacks harder.
We also consider the security of the stored information in the kernel and discuss the possibility of an attacker manipulating it.
Finally, we extract the state machines and syscall origins from several real-world applications and analyze them.
We evaluate several security-relevant metrics such as the number of states in the state machine, the number of average possible transitions per state, and the average number of allowed syscalls per syscall location.

\subsection{Performance}\label{sec:eval-performance}

\subsubsection{Setup}
All performance evaluations are performed on an i7-4790K running Ubuntu 21.04 and our modified Linux 5.13 kernel.
For all evaluations, we ensure a stable frequency. %

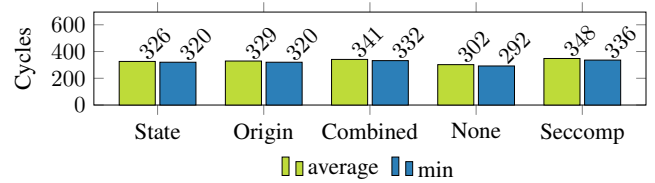
\begin{figure}[t]
  \resizebox{1\hsize}{!}{
    \begin{tikzpicture}
\begin{axis}[
ybar, ymin=0,ymax=695, ylabel=Cycles,
table/col sep=comma,
legend style={at={(0.5,-0.5)},
anchor=north,legend columns=2,draw=none},
symbolic x coords={State,Origin,Combined,None,Seccomp},
xtick=data,
nodes near coords,
every node near coord/.append style={rotate=45, anchor=west},
bar width=0.55cm,
enlarge x limits ={0.15},
width=10cm,
height=3cm,
]
\addplot[fill=green] coordinates { (State, 326) (Origin, 329) (Combined, 341) (None, 302) (Seccomp, 348)};
\addplot[fill=blue] coordinates { (State, 320) (Origin, 320) (Combined, 332) (None, 292) (Seccomp, 336)};
\legend{average \qquad, min}
\end{axis}
\end{tikzpicture}
  }
  \caption{Microbenchmark of the \textit{getppid} syscall over \SIx{100} million executions.
  We evaluate SFIP with only state machine, only syscall origin, both, and no enforcement active.
  For comparison, we also show the overhead seccomp introduces on the syscall execution.
  Latency is given in cycles.}
  \label{fig:getppid}
\end{figure}

\subsubsection{Microbenchmark}\label{sec:microbenchmark}
We perform a microbenchmark to determine the overhead our protection introduces on syscall executions.
Our benchmark evaluates the latency of the \textit{getppid} syscall, a syscall without side effects that is also used by kernel developers and previous works~\cite{Bueso2019getppid,Canella2021chestnut,Hromatka2018}.
\ToolName first extracts the state machine and the syscall-origin information from our benchmark program.
We execute the benchmark program once for every mode of SFIP, \ie state machine, syscall origins, and combined.
Each execution measures the latency of \SIx{100} million syscall invocations. %
For comparison, we also benchmark the execution with no protection active.
As with seccomp, syscalls performed while our protection is active require the slow syscall enter path to be taken due to \textit{\_TIF\_WORK\_SYSCALL\_ENTRY} being set.
This is not the case for the benchmark without protection.
As the slow path introduces part of the overhead, we additionally measure the performance of seccomp in the same experiment setup.

\paragraph{Results}
\Cref{fig:getppid} shows the results of the microbenchmark.
Our results indicate a low overhead for the syscall execution for all SFIP modes.
Transition checks show an overhead of \SI{8.15}{\percent}, syscall origin \SI{9.13}{\percent}, and combined \SI{13.1}{\percent}.
Seccomp introduces an overhead of \SI{15.23}{\percent}.
We expect that large parts of the overhead are due to the slow and complex syscall enter path instead of the checks itself.
The improved seccomp has a complexity of $\mathcal{O}(1)$ for simple allow/deny filters~\cite{Corbet2020syscall}, same as our state machine.
The syscall-origin check has a complexity of $\mathcal{O}(N)$, with typically small numbers for $N$, \ie $N=1$ for the \textit{getppid} syscall in the microbenchmark.
\cref{sec:sysloc-analysis} provides a more thorough evaluation of $N$ in real-world applications.
The additional overhead in seccomp is due to its filters being written in cBPF and converted to and executed as eBPF.

\begin{figure}[t]
  \resizebox{1\hsize}{!}{
    \begin{tikzpicture}
\begin{axis}[
ybar, ymin=0,ymax=51, ylabel=Seconds,
table/col sep=comma,
legend style={at={(0.5,-0.5)},
anchor=north,legend columns=2,draw=none},
symbolic x coords={State,Origin,Combined,None},
xtick=data,
nodes near coords,
every node near coord/.append style={rotate=45, anchor=west},
bar width=0.55cm,
enlarge x limits ={0.15},
width=10cm,
height=3cm,
]
\addplot[fill=green] coordinates { (State, 25.23) (Origin, 25.26) (Combined, 25.34) (None, 24.96)};
\addplot[fill=blue] coordinates { (State, 9.78) (Origin, 9.69) (Combined, 9.58) (None, 9.41)};
\legend{ffmpeg\qquad, nginx}
\end{axis}
\end{tikzpicture}
  }
  \caption{We perform a macrobenchmark of the two largest applications in our set of real-world applications.
  For nginx, we use the \texttt{ab} tool to make \SIx{100000} requests and time how long it takes to process them.
  For ffmpeg, we convert a video (\SI{21}{\mega\byte}) from one file format to another.
  In both cases, we perform the test \SIx{100} times for each mode of SFIP.}
  \label{fig:macro}
\end{figure}
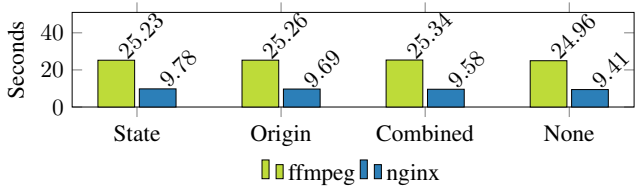

\subsubsection{Macrobenchmark}\label{sec:macrobenchmark}
To demonstrate that SFIP can be applied to large scale, real-world applications with a minimal performance overhead, we perform a macrobenchmark.
We benchmark two of the larger applications used in previous work~\cite{Canella2021chestnut,Ghavamnia2020temporal}, \ie nginx and ffmpeg.
We extract the syscall state machine and syscall-origin information for each of the two applications.
We measure the performance over \SIx{100} executions with only state machine, only syscall origin, both, and no enforcement active.
For nginx, we start the server and measure the time it takes to process \SIx{100000} requests.
For ffmpeg, we convert a video (\SI{21}{\mega\byte}) from one file format to another.
In both cases, we verified that syscalls are being executed, \eg each request for nginx executes at least \SIx{13} syscalls.

\paragraph{Results}
\cref{fig:macro} shows the results of the macrobenchmark.
In nginx, we observe a small increase in execution time when any mode of SFIP is active.
On average, we saw an increase from \SI{24.96}{\second} to \SI{25.34}{\second} in case both checks are performed.
Hence, the overhead is negligible and unnoticable for most users.
We observe similar overheads in the case of our ffmpeg benchmark. %
For the combined checks, we only observe an increase from \SI{9.41}{\second} to \SI{9.58}{\second}.
ffmpeg demonstrates one interesting result: the overhead for the combined checks is less than the overhead for each individual component.
Across several repetitions of the benchmark, this effect was always present.
Independent of the application, our results demonstrate that SFIP is a feasible concept for modern, large scale applications.

\begin{table}[t]
    \centering
    \caption{The results of our extraction time evaluation in real world applications.\
        We present both the compilation time of the respective application with and without our extraction active.\
    }
    \label{tbl:eval-extraction}
    \resizebox{\hsize}{!}{
        \begin{tabular}{lcc}
            \textbf{Application} & \textbf{\makecell{Unmodified                                                      \\Average / SEM}} & \textbf{\makecell{Modified\\Average / SEM}} \\
            \toprule
            ffmpeg               & \SI{162.12}{\second} / \SIx{0.78}           & \SI{1783.15}{\second} / \SIx{10.61} \\
            \midrule
            mupdf                & \phantom{0}\SI{58.01}{\second} / \SIx{0.71} & \phantom{0}\SI{489.85}{\second} / \SIx{0.68}   \\
            \midrule
            nginx                & \phantom{00}\SI{8.22}{\second} / \SIx{0.03} & \phantom{0}\SI{226.64}{\second} / \SIx{1.67}   \\
            \midrule
            busybox              & \phantom{0}\SI{16.09}{\second} / \SIx{0.08} & \phantom{00}\SI{81.33}{\second} / \SIx{0.14}    \\
            \midrule
            coreutils            & \phantom{00}\SI{5.50}{\second} / \SIx{0.02} & \phantom{00}\SI{14.39}{\second} / \SIx{0.41}    \\
            \bottomrule
        \end{tabular}
    }
\end{table}

\subsubsection{Extraction-Time Benchmark}\label{sec:eval-extraction}
We evaluate the time it takes to extract the information required for the state machine and syscall origins. 
As targets, we use several real-world applications (\cf \cref{tbl:eval-extraction}) used in previous works on automated seccomp sandboxing~\cite{Canella2021chestnut,Ghavamnia2020temporal,DeMarinis2020}.
These range from smaller utility applications such as busybox and coreutils to applications with a larger and more complex codebase such as ffmpeg, mupdf, and nginx.
For the benchmark, we compile each application \SIx{10} times using our modified compiler with and without our extraction active.
The resulting applications all use musl's implementation of libc and are static binaries.

\paragraph{Results}
\cref{tbl:eval-extraction} shows the result of the extraction-time benchmark.
We present the average compilation time and the standard error for compiling each application \SIx{10} times.
The results indicate that the extraction introduces a significant overhead.
For instance, in the case of the coreutils applications, we observe an increase in compilation time from approximately \SI{6}{\second} to \SI{15}{\second}.
We observe the largest increase in nginx with an increase from approximately \SI{8}{\second} to \SI{227}{\second}.
Naturally, the larger the application, the longer our extraction takes as more functions must be visited.
Most of the overhead materializes in the linker while the extraction in the frontend and backend is fast.
We expect that a full implementation can significantly improve upon the extraction time by employing more efficient caching and by potentially applying other construction algorithms.

Similar to previous work~\cite{Ghavamnia2020temporal}, we consider the increase in compilation time not to be prohibitive as it is a one-time cost when releasing a new version of the application.
Hence, the improvement in security outweighs the increase in compilation time.

\subsection{Security}\label{sec:eval-security}
In this section, we evaluate the security provided by SFIP.
We discuss the theoretical security benefit of each mode of SFIP in the context of control-flow-hijacking attacks.
We then evaluate a real vulnerability in BusyBox version 1.4.0 and later\footnote{\url{https://ssd-disclosure.com/ssd-advisory-busybox-local-cmdline-stack-buffer-overwrite/}}.
We also consider mimicry attacks~\cite{Wagner2001static,Wagner2002mimicry} and perform an analysis of real-world state machines and syscall origins.

\subsubsection{Syscall-Flow Integrity in the Context of Control-flow Hijacking}
In the threat model of SFIP (\cf \cref{sec:threat-model}), an attacker has gained control over the program-counter value of an unprivileged application.
In such a situation, an attacker can either inject code, so-called shellcode, that is then executed, or reuse existing code in a so-called code-reuse attack. 
In a shellcode attack, an attacker manages to inject their own custom code into the address space of a running application.
With control over the program-counter value, an attacker can redirect the control flow to the injected code.
On modern systems, these types of attacks are by now harder to execute due to data execution prevention~\cite{Szekeres2013sok,Microsoft2021dep}, \ie data is no longer executable.
An attacker must not only be able to inject their own code and gain control over the program counter, but also make the injected code executable.
This process of making data executable requires syscalls, \eg the \textit{mprotect} syscall.
For this, an attacker has to rely on existing code (gadgets) in the exploited application to execute such a syscall.
An attacker might be lucky, and the correct parameters are already present in the respective registers. 
Then, the control flow only has to be diverted to a code sequence that executes the syscall, resulting in a straightforward code-reuse attack commonly known as return2libc~\cite{Nergal2001ret2libc}. 
Realistically, however, an attacker first has to get the location and size of the shellcode area into the corresponding registers. 
Achieving that is possible by relying on small code gadgets executing one or more instructions which an attacker can chain together to achieve arbitrary code execution. 
Depending on the type of gadgets, such attacks are commonly known as return-oriented-programming~\cite{Shacham2007} or jump-oriented-programming attacks~\cite{Bletsch2011JOP}. 
While such attacks relying purely on existing code are Turing complete~\cite{Roemer2012ropturing}, writing the entire payload in such a way is very tedious.
Hence, attackers often use code-reuse attacks to spawn a shell or make shellcode executable~\cite{Aleph1996shellcode}. 
As these workflows are common, they are also integrated into, \eg Ropper, a tool to help with such attacks~\cite{scoding2013ropper}.

On an unprotected system, every application can execute the \textit{mprotect} syscall.
Depending on the application, the \textit{mprotect} syscall can also not be blocked by seccomp if the respective application requires it.
With SFIP, attacks that rely on \textit{mprotect} can potentially be prevented even if the application requires the syscall.
First, we consider a system where only the state machine is verified on every syscall execution.
\textit{mprotect} is a rare syscall that is mainly used in the initialization phase of an application~\cite{Ghavamnia2020temporal,Canella2021chestnut}.
Hence, we expect very few other syscalls to have a transition to it, if any.
This leaves a very small window for an attacker to execute the syscall to make the shellcode executable, \ie it is unlikely that the attempt succeeds in the presence of state machine SFIP.
Still, with only state machine checks in place, the syscall can originate from any syscall instruction within the application.

Contrary, if only the syscall origin is enforced, the \textit{mprotect} syscall is only allowed at certain syscall instructions.
Hence, an attacker needs to construct a ROP chain that sets up the necessary registers for the syscall and then returns to such a location.
In most cases, the only instance where \textit{mprotect} is allowed is within the libc \texttt{mprotect} function.
If executed from there, the syscall succeeds.
If the syscall originates from another location, the check fails, and the application is terminated.
Still, with only syscall origins being enforced, the previous syscall is not considered, allowing an attacker to perform the attack at any point in time.

With both active, \ie full SFIP, several restrictions are applied to a potential attack.
The attacker must construct a ROP chain that either starts after a syscall with a valid transition to \textit{mprotect} was executed, or the ROP chain must contain a valid sequence of syscalls that lead to such a state, \ie a mimicry attack (\cf \cref{sec:eval-mimicry}).
Additionally, all syscalls must originate from a location where they can legally occur.
The attack succeeds only in this special case.
These additional constraints significantly increase the security of the system.

\subsubsection{Real-world Exploit}
For a real-world application, we evaluate a stack-based buffer overflow present in the BusyBox arp applet from version 1.4.0 to version 1.23.1. 
In line with our threat model, we assume that all software-based security mechanisms, such as ASLR and stack protector, have already been circumvented. 
The vulnerable code is in the \texttt{arp\_getdevhw} function, which copies a user-provided command-line parameter to a stack-allocated structure using \texttt{strcpy}. 
By providing a device name longer than \texttt{IFNAMSIZ} (default 16 characters), this overflow overwrites the stack content, including the stored program counter. 

The simplest exploit we found is to mount a return2libc attack using a \textit{one gadget RCE}, \ie a gadget that directly spawns a shell. 
In libc version 2.23, we discovered such a gadget at offset \texttt{0xf0897}, with the only requirement that offset \texttt{0x70} on the stack is zero, which is luckily the case. 
Hence, by overwriting the stored program counter with that offset, we can successfully replace the application with an interactive shell. 
With SFIP, this exploit is prevented. 
Running the exploit executes the \textit{socket} syscall right before the \textit{execve} syscall that opens the shell. 
While the \textit{execve} syscall is at the correct location, the state machine does not allow a transition from the \textit{socket} to the \textit{execve} syscall. 
Hence, exploits that directly open a shell are prevented. 
We also verified that there is no possible transition from \textit{socket} to \textit{mprotect}, hence the loaded shellcode cannot be marked as executable. 
There are only \SIx{21} syscalls after a \textit{socket} syscall allowed by the state machine. 
Especially as neither the \textit{mprotect} nor the \textit{execve} syscall are available, possible exploits are drastically reduced. 
To circumvent the protection, an attacker would need to find gadgets allowing a valid transition chain from the \textit{socket} to the \textit{execve} (or \textit{mprotect}) syscall. 
We also note that the buffer overflow itself is also a limiting factor. 
As the overflow is caused by a \texttt{strcpy} function, the exploit payload, \ie the ROP chain, cannot contain any null byte. 
Thus, given that user-space addresses on 64-bit systems always have the 2 most-significant address bits set to 0, a longer chain is extremely difficult to craft. 

\subsubsection{Syscall-Flow-Integrity Protection and Mimicry Attacks}\label{sec:eval-mimicry}
We consider the possibility of mimicry attacks~\cite{Wagner2001static, Wagner2002mimicry}. %
In a mimicry attack, an attacker tries to circumvent a detection system by evading the policy.
For instance, if an intrusion-detection system is trained to detect a specific sequence of syscalls as malicious, an attacker can add arbitrary, for the attack unneeded, syscalls that hide the actual attack.
With SFIP, such attacks become significantly harder.
An attacker needs to identify the last executed syscall and knowledge of the valid transitions for all syscalls.
With this knowledge, the attacker then needs to perform a sequence of syscalls that forces the state machine into a state where the malicious syscall is a valid transition.
Additionally, as syscall origins are enforced, the attacker has to do this in a ROP attack and is limited to syscall locations where the specific syscalls are valid.
While this does not make mimicry attacks impossible, it adds several constraints that make the attack significantly harder.

\subsubsection{Security of Syscall-Flow Information in the Kernel}\label{sec:eval-info-kernel}
The security of the syscall-flow information stored in the kernel is crucial for effective enforcement.
Once the application has sent the information to the kernel for enforcement, it is the responsibility of the kernel to prevent malicious changes to the information.
The case where the initial information sent to the kernel is malicious is outside of the threat model (\cf \cref{sec:threat-model}).

The kernel stores the information in kernel memory; hence direct access and manipulation is not possible.
The only way to modify the information is through our new syscall.
Our implementation currently does not allow for any changes to the installed information, \ie no updates are allowed.
An attacker using our syscall and a ROP attack to manipulate the information is also not possible as the syscall itself needs to pass SFIP checks before being executed.
As the application contains no valid transition nor location for the syscall, the kernel terminates the application.

Still, as allowing no updates is a design decision, another implementation might consider allowing updates.
In this case, the application needs to perform our new syscall to update the filters.
Before our syscall is executed, SFIP is applied to the syscall, \ie it is verified whether there is a valid transition to it and whether it originates at the correct location.
If not, the kernel terminates the application; otherwise, the update is applied.
In this case, if timed correctly, an attacker is able to maliciously modify the stored information.

\subsubsection{State Machine Reachability Anaysis}\label{sec:eval-security:reachability}
We analyse the state machine of several real-world applications in more detail. %
We define a state in our state machine as a syscall with at least one outgoing transition.
While Wagner and Dean~\cite{Wagner2001static} only provide information on the \textit{average branching factor}, \ie the number of average transitions per state, we extend upon this to provide additional insights into automatically generated syscall state machines.
We focus on several key factors: the overall number of states in the application and the minimum, maximum, and average number of transitions across these states.
These are key factors that determine the effectiveness of SFIP.
We do not consider additional protection provided by enforcing syscall origins.
We again rely on real-world applications that have been used in previous work~\cite{Canella2021chestnut,DeMarinis2020,Ghavamnia2020temporal}.
For busybox and coreutils, we do not provide the data for every utility individually, but instead present the average of all contained utilities, \ie \SIx{398} and \SIx{103}, respectively.
To determine the improvement in security, we consider an unprotected version of the respective application, \ie every syscall can follow the previously executed syscall.
Additionally, we compare our results to a seccomp-based version. %

\begin{table}[t]
\centering
\caption{We evaluate various properties of applications station machines.\
    These metrics include the average number of transitions per state, the number of states in the state machine, min and max transitions.\
      The numbers for busybox and coreutils are the averages over all individual utilites (\SIx{398} and \SIx{103} utilities, respectively).}
\label{tbl:state-analysis}
\resizebox{\hsize}{!}{
\begin{tabular}{lcccc}
\textbf{Application} & \textbf{Average Transitions} & \textbf{\#States} & \textbf{Min Transitions} & \textbf{Max Transitions} \\
\toprule
busybox & \SIx{15.73} & \phantom{0}\SIx{24.51} & \SIx{1} & \SIx{21.09} \\
\midrule
muraster & \SIx{17.51} & \phantom{0}\SIx{41.00} & \SIx{1} & \SIx{33.00} \\
\midrule
nginx & \SIx{65.55} & \SIx{108.00} & \SIx{1} & \SIx{80.00} \\
\midrule
coreutils & \SIx{15.75} & \phantom{0}\SIx{27.11} & \SIx{1} & \SIx{23.00} \\
\midrule
ffmpeg & \SIx{48.48} & \phantom{0}\SIx{56.00} & \SIx{1} & \SIx{51.00} \\
\midrule
mutool & \SIx{32.00} & \phantom{0}\SIx{61.00} & \SIx{1} & \SIx{46.00} \\
\bottomrule
\end{tabular}
}
\end{table}

\paragraph{Results}
We show the results of this evaluation in \cref{tbl:state-analysis}.
nginx shows the highest number of states with \SIx{108}, followed by mutool and ffmpeg with \SIx{61} and \SIx{56} states, respectively.
This is to be expected as they have the largest code base and provide many different functionalities.
coreutils and busybox also provide multiple functionalities but split across various utilities.
Hence, their number of states is comparatively low.

Interestingly, each application has at least one state with only one valid transition.
We manually verified this transition, and in every case, it is a transition from the \textit{exit\_group} syscall to the \textit{exit} syscall.
Based on the source code of musl, this is indeed the only valid transition for this syscall.

The combination of the average and maximum number of transitions together with the number of states provides some interesting insight.
We observe that in most cases, the number of average transitions is relatively close to the maximum number of transitions, while the difference to the number of states can be larger.
This indicates that our state machine is heavily interconnected, which is to be expected when we consider the design of modern applications and what syscalls are used for.
Consider a normal application written in a high-level language.
The application must delegate certain tasks via syscalls to the kernel, such as allocating memory, sending data over the network, or writing to a file.
As syscalls can fail, they are often followed by error checking code that performs application-specific error handling, logs the error, or terminates the application.
As these tasks additionally require syscalls, a potential transition to these syscalls is automatically detected, leading to larger state machines.
Another source is locking, as the involved syscalls can be preceded and followed by a wide variety of other syscalls.
Additionally, the overapproximation of indirect calls also increases the number of transitions.

Even with such interconnected state machines, the security improvement is still large compared to an unprotected version of the application or even a seccomp-based version.
In the case of an unprotected version, all syscalls are valid successors to a previously executed syscall.
An unmodified Linux kernel 5.13 provides \SIx{357} syscalls.
Compared to nginx, which has the highest number of average transitions with \SIx{66}, this is an increase of factor \SIx{5.4} in terms of available transitions.
In our state machine, the number of states corresponds to the number of syscalls an automated approach needs to allow for seccomp-based protection.
These numbers also match the numbers provided in previous work on automated seccomp filter generation.
Canella~\etal\cite{Canella2021chestnut} reported \SIx{105} syscalls in nginx and \SIx{63} in ffmpeg.
Ghavamnia~\etal\cite{Ghavamnia2020temporal} reported \SIx{104} in nginx.
Each such syscall can follow any of the other syscalls that are part of the set.
In the case of nginx, this is around factor \SIx{1.6} more than in the average state when SFIP is applied.
Hence, we conclude that even coarse-grained SFIP can drastically increase the security of the system.

\begin{table*}[t]
      \centering
      \caption{We evaluate various metrics for our syscall location enforcement.\
            The metrics include the total number of functions containing syscalls, min and max and average number of syscalls per function, total syscall offsets found, average offsets per syscall, and the number of syscalls in the musl syscall wrapper functions used by the application.\
            The numbers for busybox and coreutils are the averages over all contained utilites (\SIx{398} and \SIx{103} utilities, respectively).}
      \label{tbl:sysloc-analysis}
      \resizebox{\hsize}{!}{
            \begin{tabular}{lcccccccc}
                  \textbf{Application} & \textbf{\#Functions}   & \textbf{Min Syscalls} & \textbf{Max Syscalls} & \textbf{Avg Syscalls / Function} & \textbf{Total \#Offsets} & \textbf{Avg \#Offsets} & \textbf{\#syscall()} & \textbf{\#syscall\_cp()} \\
                  \toprule
                  busybox              & \phantom{0}\SIx{30.57} & \SIx{1}               & \phantom{0}\SIx{9.83} & \SIx{1.48}                       & \SIx{102.64}             & \SIx{3.75}             & \SIx{1.71}           & \phantom{0}\SIx{9.79}               \\
                  \midrule
                  muraster             & \phantom{0}\SIx{55.00} & \SIx{1}               & \SIx{12.00}           & \SIx{1.62}                       & \SIx{193.00}             & \SIx{4.60}             & \SIx{0.00}           & \phantom{0}\SIx{4.00}                  \\
                  \midrule
                  nginx                & \SIx{105.00}           & \SIx{1}               & \SIx{24.00}           & \SIx{1.53}                       & \SIx{318.00}             & \SIx{3.00}             & \SIx{7.00}           & \SIx{24.00}                 \\
                  \midrule
                  coreutils            & \phantom{0}\SIx{36.86} & \SIx{1}               & \phantom{0}\SIx{4.21} & \SIx{1.38}                       & \SIx{116.71}             & \SIx{4.42}             & \SIx{1.00}           & \phantom{0}\SIx{3.41}               \\
                  \midrule
                  ffmpeg               & \phantom{0}\SIx{89.00} & \SIx{1}               & \SIx{13.00}           & \SIx{1.55}                       & \SIx{279.00}             & \SIx{4.98}             & \SIx{0.00}           & \SIx{13.00}                 \\
                  \midrule
                  mutool               & \phantom{0}\SIx{81.00} & \SIx{1}               & \SIx{14.00}           & \SIx{1.67}                       & \SIx{278.00}             & \SIx{4.15}             & \SIx{6.00}           & \SIx{14.00}                 \\
                  \bottomrule
            \end{tabular}
      }
\end{table*}

\subsubsection{Syscall Origins Analysis}\label{sec:sysloc-analysis}
We perform a similar analysis for our syscall origins in real-world applications.
We focus on analyzing the number of syscall locations per application, and for each such location, the number of syscalls that can be executed.
Special focus is put on the number of syscalls that can be invoked through the syscall wrapper functions as they can allow a wide variety of syscalls. %
Hence, the fewer syscalls are available through these functions, the better the security of the system.

\paragraph{Results}
We show the results of this evaluation in \cref{tbl:sysloc-analysis}.
The average number of offsets per syscall indicates that most syscalls are available at multiple locations.
This is most likely due to the inlining of the syscall.
This number is largely driven by the \textit{futex} syscall, as locking is required in many places of applications.
Error handling is a less driving factor in this case as these are predominantly printed using dedicated, non-inlined functions.

The last two columns analyze the number of syscalls that can be invoked by the respective syscall wrapper function and demonstrate a non-bijective mapping of syscalls to syscall locations.
Relatively few syscalls are available through the \texttt{syscall()} function as it can be more easily inlined, \ie it is almost always inlined within libc itself.
On the other hand, \texttt{syscall\_cp()} cannot be inlined as it is a wrapper around an aliased function that performs the actual syscall.

Our results also indicate that, on average, every function that contains a syscall contains more than one syscall.
nginx contains the most functions with a syscall and the highest number of total syscall offsets.
Hence, without syscall-origin enforcement, an attacker can choose from \SIx{318} syscall locations to execute any syscall during a ROP attack.
With our enforcement, the number is drastically reduced as each one of these locations can, on average, perform only \SIx{3} syscalls instead of \SIx{357}.

\section{Discussion}\label{sec:discussion}
\paragraph{Limitations and Future Work}
Our proof-of-concept implementation currently does not handle signals and syscalls invoked in a signal handler.
A full implementation can efficiently solve this limitation.
Wagner and Dean~\cite{Wagner2001static} propose to additionally monitor signal events and add a pre-guard and post-guard event to the control-flow graph.
With that, certain transitions in the control-flow graph are then only possible upon the reception of such an event.
Our proposal deviates from that, \ie does not require pre- and post-guard events in the control-flow graph, makes checks easier, and reduces the size of the main state machine.
The compiler can identify all functions that serve as a signal handler and the functions that are reachable through it.
Hence, it can extract a per-signal state machine to which the kernel switches when it sets up the signal stack frame.
This allows for small per-signal state machines, which further improve security.
As this requires significant engineering work, we leave the implementation and evaluation for future work.
Note that a similar approach might be feasible for generating per-thread state machines.

The state-machine construction we proposed in this paper leads to coarse-grained state machines.
We propose an improvement that leads to fine-grained state machines with improved security.
The improvement is based on the fact that we can statically identify syscall origins.
Future work can intertwine this information on a deeper level with the generated state machine.
By doing so, a transition to another state is then not only dependent on the previous and the current syscall number but also on the location of the previous and current syscall instruction in the virtual address space.
This allows to better represent the syscall-flow graph of the application without relying on context-sensitivity or call stack information~\cite{Wagner2001static,Giffin2004efficient,Sekar2001automaton}.
As this requires significant changes to the compiler and the enforcement in the kernel, as well as a thorough evaluation, we leave this for future work.

\paragraph{Related Work}
In 2001, the seminal work by Wagner and Dean~\cite{Wagner2001static} introduced automatically generated syscall NDFAs, NDPDAs, and digraphs for sequence checks in intrusion-detection systems.
We build upon the concept of a syscall digraph that they introduced but modify its construction and representation to allow for better performance. %
Our work then further extends upon theirs by additionally verifying whether a syscall originates from a valid location.
The accuracy and performance of our syscall-flow-integrity protection allows for real-time enforcement in large-scale applications.

Several papers have focused on extracting and modeling the control flow of an application, based on the work by Forrest~\etal\cite{Forrest1996self}.
Frequently, such approaches rely on dynamic analysis of repeated runs of the application~\cite{Garvey1991model,Ghosh1999profiles,Hofmeyr1998sequence,Ilgun1995state,Lane1999temporal,Wenke1999mining,Lunt1988automated,Teng1990adaptive,Wespi2000audit}.
Other approaches rely on machine-learning techniques to learn the sequence of syscalls~\cite{Lv2018prediction} or to detect intrusions~\cite{Zhengdao2008intrusion,Qiao2002hmm}.
Giffin~\etal\cite{Giffin2005environment} proposed incorporating environment information in the static analysis to generate more precise models.
Other works have disregarded the control flow and instead focused on detecting intrusions based on syscall arguments alone~\cite{Kruegel2003arguments,Mutz2006anomaly}.
Forrest~\etal\cite{Forrest2008evolution} provide an analysis on the evolution of system-call monitoring.

Two prominent approaches to learn the sequence of syscalls are VtPath~\cite{Feng2003anomaly} and the Dyck model~\cite{Giffin2004efficient}.
Both consider additional stack information and rely on context-sensitive models.
Our work differs as we do not require stack information, context-sensitive models, dynamic tracing of an application, or code instrumentation.
The only additional information we consider is the mapping of syscalls to syscall instructions.

A recent study provides a modern implementation of a syscall-sequence-based intrusion-detection system that relies on hidden markov models~\cite{Byrnes2020sequence}.
The approach relies on eBPF to hook the syscall exit handler.
Since the discovery of Spectre~\cite{Kocher2019}, kernel developers have restricted access to eBPF.
New, unprivileged use cases have faced considerable pushback from the kernel developers, making such an approach infeasible~\cite{Corbet2021seccompebpf}.

Recent, orthogonal work has investigated the possibility of automatically generating filters for seccomp~\cite{DeMarinis2020,Canella2021chestnut,Ghavamnia2020temporal,Ghavamnia2020confine}, either from the source code~\cite{Canella2021chestnut,Ghavamnia2020temporal} or from a binary~\cite{DeMarinis2020,Canella2021chestnut}.
We expect that \ToolName can be extended to generate the required information from binaries.
More recent work proposed an alternative to seccomp that maintains the speed of seccomp while also providing a secure way to perform complex argument checks~\cite{Canella2021isolation}.
In contrast to these works, we consider sequences of syscalls and the origin of a syscall, which requires additional challenges to be solved (\cf \cref{sec:challenges}).

A similar approach to our syscall-origin enforcement has been proposed by Linn~\etal\cite{Linn2005sysloc} and de Raadt~\cite{Theo2019sysorigin}.
The former extracts the syscall locations and numbers from a binary and enforces them on the kernel level, but their technique fails in the presence of ASLR.
The latter is only able to restrict the execution of syscalls to entire regions of a binary, but not a precise location, \ie the entire text segment of a static binary is a valid origin.
Additionally, in the entire region, any syscall is valid.
Our work improves upon these works in several ways:
First, we present a way to enforce the syscall location in the presence of ASLR, improving upon the former.
Second, our approach limits the execution of specific syscalls to precise locations, improving upon the latter.
Third, we improve upon both by considering the security benefits when combining such an approach with syscall state machines.

\section{Conclusion}\label{sec:conclusion}
In this paper, we introduced the concept of syscall-flow-integrity protection (SFIP), complementing the concept of CFI with integrity for user-kernel transitions.
We showed that SFIP can be implemented based on three pillars: a syscall state machine, representing possible syscall transitions; a syscall-origin mapping, which maps syscalls to the locations at which they can occur; an efficient enforcement machanism, implemented within the Linux kernel.
Based on these pillars, we demonstrated that SFIP can be fully automated on the compiler and operating-system level. 
Similar to seccomp, SFIP is opt-in, and thus fully backward compatible with legacy applications and operating systems. 
In our evaluation, we showed that SFIP can be applied to large scale applications with minimal slowdowns.
In a micro- and a macrobenchmark, we observed an overhead of only \SI{13.1}{\percent} and \SI{1.8}{\percent}, respectively.
In terms of security, we discussed and demonstrated its effectiveness in preventing control-flow-hijacking attacks in real-world applications.
Finally, to highlight the reduction in attack surface, we performed an analysis of the state machines and syscall-origin mappings of several real-world applications.
On average, we showed that SFIP decreases the number of possible transitions by \SI{38.6}{\percent} compared to seccomp and \SI{90.9}{\percent} when no protection is applied.

\ifAnon

\else
\section*{Acknowledgments}
This project has received funding from the European Research Council (ERC) under the European Union's Horizon 2020 research and innovation program (grant agreement No 681402).
Additional funding was provided by generous gifts from ARM and from Amazon.
Any opinions, findings, and conclusions or recommendations expressed in this paper are those of the authors and do not necessarily reflect the views of the funding parties.
\fi

{\bibliographystyle{IEEEtranS}
\bibliography{main}}

\begin{thebibliography}{10}
\providecommand{\url}[1]{#1}
\csname url@samestyle\endcsname
\providecommand{\newblock}{\relax}
\providecommand{\bibinfo}[2]{#2}
\providecommand{\BIBentrySTDinterwordspacing}{\spaceskip=0pt\relax}
\providecommand{\BIBentryALTinterwordstretchfactor}{4}
\providecommand{\BIBentryALTinterwordspacing}{\spaceskip=\fontdimen2\font plus
\BIBentryALTinterwordstretchfactor\fontdimen3\font minus
  \fontdimen4\font\relax}
\providecommand{\BIBforeignlanguage}[2]{{%
\expandafter\ifx\csname l@#1\endcsname\relax
\typeout{** WARNING: IEEEtranS.bst: No hyphenation pattern has been}%
\typeout{** loaded for the language `#1'. Using the pattern for}%
\typeout{** the default language instead.}%
\else
\language=\csname l@#1\endcsname
\fi
#2}}
\providecommand{\BIBdecl}{\relax}
\BIBdecl

\bibitem{Abadi2005CFI}
M.~Abadi, M.~Budiu, U.~Erlingsson, and J.~Ligatti, ``{Control-Flow
  Integrity},'' in \emph{{CCS}}, 2005.

\bibitem{Andersen1994points}
L.~O. Andersen, ``{Program Analysis and Specialization for the C Programming
  Language},'' Ph.D. dissertation, 1994.

\bibitem{AndroidAppSandbox}
\BIBentryALTinterwordspacing
Android, ``{Application Sandbox},'' 2021. [Online]. Available:
  \url{https://source.android.com/security/app-sandbox}
\BIBentrySTDinterwordspacing

\bibitem{AppArmor}
\BIBentryALTinterwordspacing
AppArmor, ``{AppArmor: Linux kernel security module},'' 2021. [Online].
  Available: \url{https://apparmor.net/}
\BIBentrySTDinterwordspacing

\bibitem{Bletsch2011JOP}
T.~K. Bletsch, X.~Jiang, V.~W. Freeh, and Z.~Liang, ``{Jump-oriented
  programming: a new class of code-reuse attack},'' in \emph{{AsiaCCS}}, 2011.

\bibitem{Bueso2019getppid}
\BIBentryALTinterwordspacing
D.~Bueso, ``{tools/perf-bench: Add basic syscall benchmark},'' 2019. [Online].
  Available: \url{https://lore.kernel.org/patchwork/patch/1048777/}
\BIBentrySTDinterwordspacing

\bibitem{Burow2017cfi}
N.~Burow, S.~A. Carr, J.~Nash, P.~Larsen, M.~Franz, S.~Brunthaler, and
  M.~Payer, ``{Control-Flow Integrity: Precision, Security, and Performance},''
  \emph{{ACM Computing Surveys}}, 2017.

\bibitem{Byrnes2020sequence}
{Byrnes, Jeffrey and Hoang, Thomas and Mehta, Nihal Nitin and Cheng, Yuan},
  ``{A Modern Implementation of System Call Sequence Based Host-based Intrusion
  Detection Systems},'' in \emph{{TPS-ISA}}, 2020.

\bibitem{Canella2021chestnut}
C.~Canella, M.~Werner, D.~Gruss, and M.~Schwarz, ``{Automating Seccomp Filter
  Generation for Linux Applications},'' in \emph{{CCSW}}, 2021.

\bibitem{Canella2021isolation}
{Canella, Claudio and Kogler, Andreas and Giner, Lukas and Gruss, Daniel and
  Schwarz, Michael}, ``{Domain Page-Table Isolation},''
  \emph{arXiv:2111.10876}, 2021.

\bibitem{Checkoway2010JOP}
S.~Checkoway, L.~Davi, A.~Dmitrienko, A.~Sadeghi, H.~Shacham, and M.~Winandy,
  ``Return-oriented programming without returns,'' in \emph{{CCS}}, 2010.

\bibitem{Corbet2020syscall}
J.~Corbet, ``{Constant-action bitmaps for seccomp()},'' 2020.

\bibitem{Corbet2021seccompebpf}
\BIBentryALTinterwordspacing
------, ``{eBPF seccomp() filters},'' 2021. [Online]. Available:
  \url{https://lwn.net/Articles/857228/}
\BIBentrySTDinterwordspacing

\bibitem{Cowan1998}
C.~Cowan, C.~Pu, D.~Maier, J.~Walpole, P.~Bakke, S.~Beattie, A.~Grier,
  P.~Wagle, Q.~Zhang, and H.~Hinton, ``Stackguard: Automatic adaptive detection
  and prevention of buffer-overflow attacks.'' in \emph{USENIX Security}, 1998.

\bibitem{Davi2014coarsecfi}
{Davi, Lucas and Sadeghi, Ahmad-Reza and Lehmann, Daniel and Monrose, Fabian},
  ``Stitching the gadgets: On the ineffectiveness of coarse-grained
  control-flow integrity protection,'' in \emph{{USENIX Security Symposium}},
  August 2014.

\bibitem{Theo2019sysorigin}
\BIBentryALTinterwordspacing
T.~de~Raadt, ``{syscall call-from verification},'' 2019. [Online]. Available:
  \url{https://lwn.net/Articles/806863/}
\BIBentrySTDinterwordspacing

\bibitem{DeMarinis2020}
N.~DeMarinis, K.~Williams-King, D.~Jin, R.~Fonseca, and V.~P. Kemerlis,
  ``{sysfilter: Automated System Call Filtering for Commodity Software},'' in
  \emph{RAID}, 2020.

\bibitem{Edge2012nonewprivs}
\BIBentryALTinterwordspacing
J.~Edge, ``{System call filtering and no\_new\_privs},'' 2012. [Online].
  Available: \url{https://lwn.net/Articles/475678/}
\BIBentrySTDinterwordspacing

\bibitem{Edge2015seccomp}
\BIBentryALTinterwordspacing
------, ``{A seccomp overview},'' 2015. [Online]. Available:
  \url{https://lwn.net/Articles/656307/}
\BIBentrySTDinterwordspacing

\bibitem{Feng2003anomaly}
{Feng, H.H. and Kolesnikov, O.M. and Fogla, P. and Lee, W. and Weibo Gong},
  ``{Anomaly detection using call stack information},'' in \emph{{S\&P}}, 2003.

\bibitem{Forrest1996self}
{Forrest, S. and Hofmeyr, S.A. and Somayaji, A. and Longstaff, T.A.}, ``{A
  sense of self for Unix processes},'' in \emph{{S\&P}}, 1996.

\bibitem{Forrest2008evolution}
{Forrest, Stephanie and Hofmeyr, Steven and Somayaji, Anil}, ``{The Evolution
  of System-Call Monitoring},'' in \emph{{ACSAC}}, 2008.

\bibitem{Garvey1991model}
{Garvey, Thomas D. and Lunt, Teresa F.}, ``{Model-based intrusion detection},''
  in \emph{{NCSC}}, 1991.

\bibitem{Ge2016kernel}
{Ge, Xinyang and Talele, Nirupama and Payer, Mathias and Jaeger, Trent},
  ``{Fine-Grained Control-Flow Integrity for Kernel Software},'' in \emph{{Euro
  S\&P}}, 2016.

\bibitem{Ghavamnia2020temporal}
S.~Ghavamnia, T.~Palit, S.~Mishra, and M.~Polychronakis, ``{Temporal System
  Call Specialization for Attack Surface Reduction},'' in \emph{{USENIX
  Security Symposium}}, 2020.

\bibitem{Ghavamnia2020confine}
{Ghavamnia, Seyedhamed and Palit, Tapti and Mishra, Shachee and Polychronakis,
  Michalis}, ``{Confine: Automated System Call Policy Generation for Container
  Attack Surface Reduction},'' in \emph{{RAID}}, 2020.

\bibitem{Ghosh1999profiles}
{Ghosh, Anup and Schwartzbard, Aaron and Schatz, Michael}, ``{Learning Program
  Behavior Profiles for Intrusion Detection},'' in \emph{{ID}}, 1999.

\bibitem{Giffin2005environment}
{Giffin, Jonathon and Dagon, David and Jha, Somesh and Lee, Wenke and Miller,
  Barton}, ``{Environment-Sensitive Intrusion Detection},'' in \emph{{RAID}},
  {2005}.

\bibitem{Giffin2004efficient}
{Giffin, Jonathon T and Jha, Somesh and Miller, Barton P}, ``{Efficient
  Context-Sensitive Intrusion Detection.}'' in \emph{{NDSS}}, 2004.

\bibitem{Goktas2014COP}
E.~G{\"{o}}ktas, E.~Athanasopoulos, H.~Bos, and G.~Portokalidis, ``Out of
  control: Overcoming control-flow integrity,'' in \emph{{S\&P}}, 2014.

\bibitem{Hind2001pointer}
M.~Hind, ``Pointer analysis: Haven't we solved this problem yet?'' in
  \emph{PASTE}, 2001.

\bibitem{Hofmeyr1998sequence}
{Hofmeyr, Steven A. and Forrest, Stephanie and Somayaji, Anil}, ``{Intrusion
  Detection Using Sequences of System Calls},'' \emph{J. Comput. Secur.}, 1998.

\bibitem{Hromatka2018}
T.~Hromatka, ``{seccomp and libseccomp performance improvements},'' 2018.

\bibitem{Ilgun1995state}
{Ilgun, K. and Kemmerer, R.A. and Porras, P.A.}, ``{State transition analysis:
  a rule-based intrusion detection approach},'' \emph{{TSE}}, 1995.

\bibitem{Android2017seccomp}
\BIBentryALTinterwordspacing
G.~Inc., ``{Seccomp filter in Android O},'' 2017. [Online]. Available:
  \url{https://android-developers.googleblog.com/2017/07/seccomp-filter-in-android-o.html}
\BIBentrySTDinterwordspacing

\bibitem{Ispoglou2018}
K.~K. Ispoglou, B.~AlBassam, T.~Jaeger, and M.~Payer, ``{Block Oriented
  Programming: Automating Data-Only Attacks},'' in \emph{CCS}, 2018.

\bibitem{Kemerlis2015}
V.~Kemerlis, ``Protecting commodity operating systems through strong kernel
  isolation,'' Ph.D. dissertation, Columbia University, 2015.

\bibitem{Kemerlis2014}
V.~P. Kemerlis, M.~Polychronakis, and A.~D. Keromytis, ``ret2dir: Rethinking
  kernel isolation,'' in \emph{USENIX Security Symposium}, 2014.

\bibitem{Kemerlis2012}
V.~P. Kemerlis, G.~Portokalidis, and A.~D. Keromytis, ``kguard: Lightweight
  kernel protection against return-to-user attacks,'' in \emph{{USENIX}
  Security Symposium}, 2012.

\bibitem{Kemmerer2002intrusion}
{Kemmerer, Richard A and Vigna, Giovanni}, ``{Intrusion detection: a brief
  history and overview},'' \emph{{Computer}}, 2002.

\bibitem{Kim2014}
Y.~Kim, R.~Daly, J.~Kim, C.~Fallin, J.~H. Lee, D.~Lee, C.~Wilkerson, K.~Lai,
  and O.~Mutlu, ``{Flipping Bits in Memory Without Accessing Them: An
  Experimental Study of DRAM Disturbance Errors},'' in \emph{ISCA}, 2014.

\bibitem{Kocher2019}
P.~Kocher, J.~Horn, A.~Fogh, D.~Genkin, D.~Gruss, W.~Haas, M.~Hamburg, M.~Lipp,
  S.~Mangard, T.~Prescher, M.~Schwarz, and Y.~Yarom, ``{Spectre Attacks:
  Exploiting Speculative Execution},'' in \emph{S\&P}, 2019.

\bibitem{Kruegel2003arguments}
{Kruegel, Christopher and Mutz, Darren and Valeur, Fredrik and Vigna,
  Giovanni}, ``{On the Detection of Anomalous System Call Arguments},'' in
  \emph{{ESORICS}}, 2003.

\bibitem{Lan2015loop}
B.~Lan, Y.~Li, H.~Sun, C.~Su, Y.~Liu, and Q.~Zeng, ``Loop-oriented programming:
  a new code reuse attack to bypass modern defenses,'' in \emph{IEEE
  Trustcom/BigDataSE/ISPA}, 2015.

\bibitem{Lane1999temporal}
{Lane, Terran and Brodley, Carla E.}, ``{Temporal Sequence Learning and Data
  Reduction for Anomaly Detection},'' \emph{{TOPS}}, 1999.

\bibitem{Linn2005sysloc}
C.~M. Linn, M.~Rajagopalan, S.~Baker, C.~Collberg, S.~K. Debray, and J.~H.
  Hartman, ``{Protecting Against Unexpected System Calls},'' in \emph{{USENIX
  Security Symposium}}, 2005.

\bibitem{Lipp2018meltdown}
M.~Lipp, M.~Schwarz, D.~Gruss, T.~Prescher, W.~Haas, A.~Fogh, J.~Horn,
  S.~Mangard, P.~Kocher, D.~Genkin, Y.~Yarom, and M.~Hamburg, ``{Meltdown:
  Reading Kernel Memory from User Space},'' in \emph{USENIX Security
  Symposium}, 2018.

\bibitem{Lunt1988automated}
{Lunt, Teresa F.}, ``{Automated Audit Trail Analysis and Intrusion Detection: A
  Survey},'' in \emph{{NCSC}}, 1988.

\bibitem{Lv2018prediction}
{Lv, Shaohua and Wang, Jian and Yang, Yinqi and Liu, Jiqiang}, ``{Intrusion
  Prediction With System-Call Sequence-to-Sequence Model},'' \emph{{IEEE
  Access}}, 2018.

\bibitem{Microsoft2021dep}
\BIBentryALTinterwordspacing
Microsoft, ``{Data Execution Prevention},'' 2021. [Online]. Available:
  \url{https://docs.microsoft.com/en-us/windows/win32/memory/data-execution-prevention}
\BIBentrySTDinterwordspacing

\bibitem{Mutz2006anomaly}
{Mutz, Darren and Valeur, Fredrik and Vigna, Giovanni and Kruegel,
  Christopher}, ``{Anomalous System Call Detection},'' \emph{{TOPS}}, 2006.

\bibitem{Nergal2001ret2libc}
Nergal, ``{The advanced return-into-lib(c) explits: PaX case study},'' 2001.

\bibitem{Aleph1996shellcode}
{One, Aleph}, ``{Smashing the Stack for Fun and Profit},'' \emph{{Phrack}},
  1996.

\bibitem{Qiao2002hmm}
{Qiao, Y. and Xin, X.W. and Bin, Y. and Ge, S.}, ``{Anomaly intrusion detection
  method based on HMM},'' \emph{{Electronics Letters}}, {2002}.

\bibitem{Reis2019SiteIsolation}
C.~Reis, A.~Moshchuk, and N.~Oskov, ``{Site Isolation: Process Separation for
  Web Sites within the Browser},'' in \emph{USENIX Security Symposium}, 2019.

\bibitem{Roemer2012ropturing}
{Roemer, Ryan and Buchanan, Erik and Shacham, Hovav and Savage, Stefan},
  ``{Return-Oriented Programming: Systems, Languages, and Applications},''
  \emph{{TISSEC}}, 2012.

\bibitem{Rogowski2017}
R.~Rogowski, M.~Morton, F.~Li, F.~Monrose, K.~Z. Snow, and M.~Polychronakis,
  ``{Revisiting Browser Security in the Modern Era: New Data-Only Attacks and
  Defenses},'' in \emph{{EuroS\&P}}, 2017.

\bibitem{Schuster2015COOP}
F.~Schuster, T.~Tendyck, C.~Liebchen, L.~Davi, A.~Sadeghi, and T.~Holz,
  ``{Counterfeit Object-oriented Programming: On the Difficulty of Preventing
  Code Reuse Attacks in {C++} Applications},'' in \emph{{S\&P}}, 2015.

\bibitem{Schwarz2019ZL}
M.~Schwarz, M.~Lipp, D.~Moghimi, J.~Van~Bulck, J.~Stecklina, T.~Prescher, and
  D.~Gruss, ``{ZombieLoad: Cross-Privilege-Boundary Data Sampling},'' in
  \emph{CCS}, 2019.

\bibitem{scoding2013ropper}
\BIBentryALTinterwordspacing
scoding.de, ``{Ropper - rop gadget finder and binary information tool},'' 2013.
  [Online]. Available: \url{https://scoding.de/ropper/}
\BIBentrySTDinterwordspacing

\bibitem{Sekar2001automaton}
{Sekar, R. and Bendre, M. and Dhurjati, D. and Bollineni, P.}, ``{A fast
  automaton-based method for detecting anomalous program behaviors},'' in
  \emph{{S\&P}}, 2001.

\bibitem{Shacham2007}
H.~Shacham, ``{The geometry of innocent flesh on the bone: Return-into-libc
  without function calls (on the x86)},'' in \emph{CCS}, 2007.

\bibitem{Spengler2013uderef}
{Spengler, Brad}, ``{Recent ARM Security Improvements},'' {2013}.

\bibitem{Szekeres2013sok}
L.~Szekeres, M.~Payer, T.~Wei, and D.~Song, ``{SoK: Eternal War in Memory},''
  in \emph{S\&P}, 2013.

\bibitem{Teng1990adaptive}
{Teng, H.S. and Chen, K. and Lu, S.C.}, ``{Adaptive real-time anomaly detection
  using inductively generated sequential patterns},'' in \emph{{S\&P}}, 1990.

\bibitem{VanSchaik2019RIDL}
S.~van Schaik, A.~Milburn, S.~Österlund, P.~Frigo, G.~Maisuradze, K.~Razavi,
  H.~Bos, and C.~Giuffrida, ``{RIDL: Rogue In-flight Data Load},'' in
  \emph{{S\&P}}, 2019.

\bibitem{Wagner2001static}
{Wagner, D. and Dean, R.}, ``{Intrusion detection via static analysis},'' in
  \emph{{S\&P}}, 2001.

\bibitem{Wagner2002mimicry}
{Wagner, David and Soto, Paolo}, ``{Mimicry Attacks on Host-Based Intrusion
  Detection Systems},'' in \emph{CCS}, 2002.

\bibitem{Wenke1999mining}
{Wenke, Lee and Stolfo, S.J. and Mok, K.W.}, ``{A data mining framework for
  building intrusion detection models},'' in \emph{{S\&P}}, 1999.

\bibitem{Wespi2000audit}
{Wespi, Andreas and Dacier, Marc and Debar, Herv\'{e}}, ``{Intrusion Detection
  Using Variable-Length Audit Trail Patterns},'' in \emph{{RAID}}, 2000.

\bibitem{Firefox2019fission}
\BIBentryALTinterwordspacing
M.~Wiki, ``{Project Fission},'' 2019. [Online]. Available:
  \url{https://wiki.mozilla.org/Project_Fission}
\BIBentrySTDinterwordspacing

\bibitem{Firefox2019sandbox}
\BIBentryALTinterwordspacing
------, ``{Security/Sandbox},'' 2019. [Online]. Available:
  \url{https://wiki.mozilla.org/Security/Sandbox}
\BIBentrySTDinterwordspacing

\bibitem{SELinuxFAQ}
\BIBentryALTinterwordspacing
S.~Wiki, ``{FAQ --- SELinux Wiki},'' 2009. [Online]. Available:
  \url{http://selinuxproject.org/w/?title=FAQ&oldid=729}
\BIBentrySTDinterwordspacing

\bibitem{Yarom2014Flush}
Y.~Yarom and K.~Falkner, ``{Flush+Reload: a High Resolution, Low Noise, L3
  Cache Side-Channel Attack},'' in \emph{USENIX Security Symposium}, 2014.

\bibitem{Zhengdao2008intrusion}
{Zhengdao, Zhang and Zhumiao, Peng and Zhiping, Zhou}, ``{The Study of
  Intrusion Prediction Based on HsMM},'' in \emph{{APSCC}}, 2008.

\end{thebibliography}

\end{document}